\documentclass[british,cits]{PoS}

\usepackage{amsmath}
\allowdisplaybreaks[4]
\usepackage{mathtools}
\usepackage[separate-uncertainty]{siunitx}
\usepackage{subfig}
\usepackage{mciteplus}
\mciteSetMidEndSepPunct{;\newline}{.}{\relax}

\newcommand{\order}[1]{\ensuremath{\mathcal{O}(#1)}}
\newcommand{\AQg}[1]{\ensuremath{A_{Qg}^{(#1)}}} 
\newcommand{\AggQ}[1]{\ensuremath{A_{gg,Q}^{(#1)}}} 
\newcommand{\AgqQ}[1]{\ensuremath{A_{gq,Q}^{(#1)}}} 
\newcommand{\AqgQ}[1]{\ensuremath{A_{qg,Q}^{(#1)}}} 
\newcommand{\AqqQPS}[1]{\ensuremath{A_{qq,Q}^{\mathrm{PS},(#1)}}} 
\newcommand{\AqqQNS}[1]{\ensuremath{A_{qq,Q}^{\mathrm{NS},(#1)}}} 
\newcommand{\AqqQTR}[1]{\ensuremath{A_{qq,Q}^{\mathrm{NS,TR},(#1)}}} 
\newcommand{\AQqPS}[1]{\ensuremath{A_{Qq}^{\mathrm{PS},(#1)}}}
\newcommand{\progname}[1]{\texttt{#1}} 

\title{{\footnotesize DESY 16-210,~~DO-TH 16/29,~~TTK-16-46}\\
       Heavy flavour corrections to polarised and unpolarised 
       deep-inelastic scattering at 3-loop order\thanks{This work was supported
       in part by the Austrian Science Fund (FWF) grant SFB F50 (F5009-N15),
       the European Commission through contract PITN-GA-2012-316704
       ({HIGGSTOOLS}) and by FP7 ERC Starting Grant  257638 PAGAP.}}

\ShortTitle{Heavy flavour corrections to DIS at 3-loop order}

\author{J.~Ablinger$^a$, \speaker{A.~Behring}$^{b,c}$,
        J.~Bl\"umlein$^b$, A.~De~Freitas$^b$, A.~Hasselhuhn$^d$,
        A.~von~Manteuffel$^{e,f}$, M.~Round$^a$, C.~Schneider$^a$,
        F.~Wi\ss{}brock$^g$\\
        \llap{$^a$}Research Institute for Symbolic Computation (RISC),
                   Johannes Kepler University,\\
                   Altenbergerstra\ss{}e 69, A--4040, Linz, Austria \\
        \llap{$^b$}Deutsches Elektronen-Synchrotron, DESY, Platanenallee 6,
                   D--15738 Zeuthen, Germany \\
        \llap{$^c$}Institut f\"ur Theoretische Teilchenphysik und Kosmologie,
                   RWTH Aachen University, \\ D--52056 Aachen, Germany \\
        \llap{$^d$}Institut f\"ur Theoretische Teilchenphysik,
                   Karlsruher Institut f\"ur Technologie (KIT),\\
                   D--76128 Karlsruhe, Germany \\
        \llap{$^e$}PRISMA Cluster of Excellence, Institute of Physics,
                   J. Gutenberg University, D--55099 Mainz,\\
                   Germany \\
        \llap{$^f$}Department of Physics and Astronomy,
                   Michigan State University, East Lansing, MI 48824, USA \\
        \llap{$^g$}IHES, 35 Route de Chartres, F--91440 Bures-sur-Yvette,
                   France \\
        E-mail: \email{arnd.behring@desy.de}}

\abstract{We report on progress in the calculation of 3-loop corrections to the
deep-inelastic structure functions from massive quarks in the asymptotic region
of large momentum transfer $Q^2$. Recently completed results allow us to obtain
the $O(a_s^3)$ contributions to several heavy flavour Wilson coefficients which
enter both polarised and unpolarised structure functions for lepton-nucleon
scattering. In particular, we obtain the non-singlet contributions to the
unpolarised structure functions $F_2(x,Q^2)$ and $x F_3(x,Q^2)$ and the
polarised structure function $g_1(x,Q^2)$. From these results we also obtain the
heavy flavour contributions to the Gross-Llewellyn-Smith and the Bjorken sum
rules.}

\FullConference{QCD Evolution 2016\\
         May 30-June 03, 2016\\
         National Institute for Subatomic Physics (Nikhef) in Amsterdam}

\begin{document}


\section{Introduction}
Deep-inelastic scattering provides a valuable way to both test the theory of
quantum chromodynamics (QCD) and to extract theory parameters from experiments.
Among these are in particular the strong coupling constant $\alpha_s$
\cite{Bethke:2011tr,*Moch:2014tta,*Alekhin:2016evh}, the parton distribution
functions (PDFs) \cite{Alekhin:2013nda,Jimenez-Delgado:2014twa,*Dulat:2015mca,
*Harland-Lang:2014zoa,*Ball:2014uwa,*Abramowicz:2015mha,Accardi:2016ndt}
and the masses of the charm and bottom quarks \cite{Alekhin:2012vu,
Alekhin:2016uxn}.
To harness the full potential of the experimental data, it is necessary to have
predictions at hand for which the theoretical uncertainties keep up with the
experimental accuracy.
Currently, the corrections from massive quarks are still missing for a complete
next-to-next-to-leading order (NNLO) analysis of the deep-inelastic scattering
World data. They can be calculated analytically at NNLO in the kinematic limit
$Q^2 \gg m^2$ \cite{Buza:1995ie}, where $Q^2$ is the virtuality of the
electro-weak gauge boson and $m$ is the mass of the heavy quark. In this paper,
we report on progress in the calculation of these heavy flavour corrections.

In Section \ref{sec:framework} we describe the framework for the heavy flavour
corrections to deep-inelastic scattering in the limit $Q^2 \gg m^2$.
In this limit, the massive operator matrix elements of the light-cone
operators are the key quantities which have not been completely computed yet.
Thus, we sketch the steps involved in their calculation.
Section \ref{sec:structure-functions} contains several applications of the
results which have been obtained so far. In particular, we illustrate the impact of the heavy
flavour Wilson coefficients on the structure function $F_2(x,Q^2)$,
$g_1(x,Q^2)$ and $x F_3(x,Q^2)$, as well as their consequences for the
polarised Bjorken sum rule and the Gross-Llewellyn-Smith sum rule.
Finally, we comment on our results for the massive operator matrix element of
the non-singlet operator for transversity in Section \ref{sec:transversity}
and conclude in Section \ref{sec:conclusions}


\section{Framework of calculation}\label{sec:framework}
The structure functions of deep-inelastic scattering can in general be written
as convolutions of PDFs and Wilson coefficients (cf., e.g., \cite{Buras:1979yt,
*Reya:1979zk,*Blumlein:2012bf}). The Wilson coefficients carry the
process-dependent information about the particular scattering process at hand
and can be calculated in perturbation theory. They receive contributions from
massless quarks and gluons as well as from massive quarks. In the following, we
will refer to the contributions of a single massive quark species (i.e. charm or
bottom quarks) as the heavy flavour contributions. Starting at 3-loop order,
there are also contributions from diagrams with two different massive quarks.
Their calculation poses quite different challenges and is discussed elsewhere
\cite{Ablinger:2011pb,*Ablinger:2012qj,*DESY-14-019}. The heavy flavour
contributions to the structure function $F_2(x,Q^2)$, for example, can be
written as \cite{Buza:1995ie,Bierenbaum:2009mv,Behring:2014eya}
\begin{align}
  \MoveEqLeft{F_2^\text{h}(x,N_F+1,Q^2,m^2) =}
  \nonumber \\ &
    x \Biggl\{
      \sum_{k=1}^{N_F} e_k^2 \Biggl[
        L_{q,2}^\text{NS}\left(x,N_F+1,\frac{Q^2}{\mu^2},\frac{m^2}{\mu^2}\right)
          \otimes \Bigl[f_k(x,\mu^2,N_F)+\bar{f}_k(x,\mu^2,N_F)\Bigr]
  \nonumber \\ &
        +\frac{1}{N_F} L_{q,2}^\text{PS}\left(x,N_F+1,\frac{Q^2}{\mu^2},\frac{m^2}{\mu^2}\right) 
          \otimes \Sigma(x,\mu^2,N_F)
  \nonumber \\ &
        +\frac{1}{N_F} L_{g,2}^\text{S}\left(x,N_F+1,\frac{Q^2}{\mu^2},\frac{m^2}{\mu^2}\right)
          \otimes G(x,\mu^2,N_F) 
      \Biggr]
  \nonumber \\ &
      + e_Q^2 \Biggl[
        H_{q,2}^\text{PS}\left(x,N_F+1,\frac{Q^2}{\mu^2},\frac{m^2}{\mu^2}\right) 
          \otimes \Sigma(x,\mu^2,N_F)
  \nonumber \\ &
        +H_{g,2}^\text{S}\left(x,N_F+1,\frac{Q^2}{\mu^2},\frac{m^2}{\mu^2}\right)
          \otimes G(x,\mu^2,N_F)
      \Biggr]
    \Biggr\}
  \label{eq:F2-wilson-coeff}
  \,.
\end{align}
Here, $f_k$, $\bar{f}_k$ and $G$ refer to the PDFs for quarks and anti-quarks of
flavour $k$ and the gluon PDF, respectively. The singlet PDF combination is
defined as $\Sigma = \sum_{k=1}^{N_F} (f_k + \bar{f}_k)$ and $\mu^2$ is the
factorisation scale. The number of light quark flavours is denoted by $N_F$ and
$e_k$ and $e_Q$ represent the charges of the light and heavy quarks.
The heavy flavour Wilson coefficients are denoted by $L_{i,a}$ and $H_{i,a}$,
where the $L$ and $H$ distinguish the cases where the electro-weak gauge boson
couples to a light or a heavy quark, respectively, and the subscripts $i$ and
$a$ label the initial state parton ($q,g$) and the structure function under
consideration. The symbol $\otimes$ represents the Mellin convolution, which
turns into a simple product of moments if we apply to it a Mellin transformation
\begin{align}
  M[f](N) &= \int_0^1 \mathrm{d}x \, x^{N-1} f(x)
  \label{eq:mellin-transformation}
  \,.
\end{align}
This introduces the Mellin variable $N$, to which we will refer in several
places in the following.

The massless Wilson coefficients have been calculated up to 3-loop order
\cite{Vermaseren:2005qc}, while the massive ones are available only semi-numerically
up to 2-loop order \cite{Laenen:1992zk,*Laenen:1992xs,*Riemersma:1994hv}.%
\footnote{For a precise implementation in Mellin space see
\cite{Alekhin:2003ev}.}
However, it was observed in \cite{Buza:1995ie} that the heavy-flavour Wilson
coefficients factorise into the massless Wilson coefficients and massive
operator matrix elements (OME) in the kinematic limit where $Q^2 \gg m^2$. In
this limit, the heavy flavour Wilson coefficients
$\mathcal{C}^\text{asymp}_{i,a}$ can be written schematically as
\cite{Buza:1995ie}
\begin{align}
  \mathcal{C}^\text{asymp}_{i,a}\left(x,N_F+1,\frac{Q^2}{\mu^2},
              \frac{m^2}{\mu^2}\right) =
    \sum_j C_{j,a}\left(x,N_F+1,\frac{Q^2}{\mu^2}\right) \otimes
      A_{ij,Q}\left(x,N_F+1,\frac{m^2}{\mu^2}\right)
    + \mathcal{O}\left(\frac{m^2}{Q^2}\right)
  \,.
\end{align}
The massless Wilson coefficients $C_{i,a}$ are evaluated for $N_F+1$ massless
flavours and the massive OMEs $A_{ij,Q}$ are calculated with $N_F$ massless and
one massive quark. More details on the formalism can be found in
\cite{Buza:1995ie,Buza:1996wv,Bierenbaum:2009mv,Behring:2014eya}. Below, we will
frequently refer to the expansion coefficients of the OMEs in powers of
$a_s=\tfrac{\alpha_s}{4\pi}$, where the coefficient of $a_s^k$ is denoted by
$A_{ij,Q}^{(k)}$.

The massive OMEs are known analytically up to 2-loop order \cite{Buza:1995ie,
Buza:1996xr,Buza:1996wv,Bierenbaum:2007qe,*Bierenbaum:2008yu,*Bierenbaum:2009zt}
including linear terms in the dimensional regulator $\varepsilon=D-4$, where $D$
is the dimensionality of space-time in dimensional regularisation
\cite{tHooft:1972fi,*Ashmore:1972uj,*Cicuta:1972jf,*Bollini:1972ui}. The extension
of these results to 3-loop order is the topic of our project and here we report on
progress in this regard.

The massive OMEs can be extracted from calculating two-point functions with
external on-shell partons and operators, which introduce additional Feynman
rules beyond those of QCD. Our calculation follows a diagrammatic approach,
where all relevant diagrams are generated using \progname{QGRAF}
\cite{Nogueira:1991ex}.
After inserting the Feynman rules, we simplify the colour and Dirac algebra using
\progname{FORM} \cite{Vermaseren:2000nd,*Tentyukov:2007mu,*Kuipers:2012rf} and
\progname{color.h} \cite{vanRitbergen:1998pn}. In this way, we express the
diagrams in terms of roughly $10^5$ scalar loop integrals. In order to compute
those, we first reduce them to a smaller number of master integrals using
integration-by-parts relations \cite{Lagrange:1760,*Gauss:1813,*Green:1828,
*Ostrogradski:1831,Chetyrkin:1980pr,*Chetyrkin:1981qh,*Tkachov:1981wb}. For this
task we use the program \progname{Reduze~2} \cite{Studerus:2009ye,
*vonManteuffel:2012np}.\footnote{\progname{Reduze~2} uses the packages
\progname{GiNaC} \cite{Bauer:2000cp} and \progname{Fermat} \cite{Lewis:Fermat}.}
A major task is then to actually calculate the master integrals. Over the years
a number of techniques have proven to be very useful for this task:
\begin{itemize}
  \item \emph{Higher hypergeometric functions} \cite{Gauss:1812,*Kummer:1836,
        *Bailey:1935,*Slater:1966,*Gradstein:1981,*Appell:1926,*Appell:1925,
        *KampeDeFeriet:1937,*Exton:1976yx,*Exton:1978,*Srivastava:1985}:
        After deriving a Feynman parameter representation for the loop
        integrals, they can sometimes be brought in a form which can be
        integrated in terms of generalised hypergeometric (${}_pF_q$) or Appell
        functions that have convergent series representations. If this is the
        case, we can expand them in $\varepsilon$ and the resulting sum
        representations can be simplified using the summation algorithms
        \cite{Schneider:2001,Karr:1981,*Schneider:2004den,*Schneider:2005rec,
        *Schneider:2005pol,*Schneider:2007sim,*Schneider:2010cla,
        *Schneider:2010alg,*Schneider:2015lnc,*Schneider:2008jsc,
        *Schneider:2016jsc,*Schneider:2016two} implemented in \progname{Sigma}
        \cite{Schneider:2001,Schneider:2007,Schneider:2013book},
        \progname{EvaluateMultiSums} and \progname{SumProduction}
        \cite{Ablinger:2010pb,*Blumlein:2012hg,*Schneider:2013,
        *Schneider:2013zna} with support from \progname{HarmonicSums}
        \cite{Ablinger:2010kw,Ablinger:2013hcp,Ablinger:2011te,Ablinger:2013cf,
        Ablinger:2014bra,Ablinger:2014rba} for dealing with the nested sums that
        arise.
  \item \emph{Mellin-Barnes integrals} \cite{Mellin:1895,*Barnes:1908,
        *Barnes:1910,Smirnov:1999gc,*Tausk:1999vh}:
        Even if a direct evaluation of the Feynman parameter integrals in terms
        of hypergeometric functions is not possible, it may still be feasible to
        derive a sum representation that can be simplified using
        \progname{Sigma} and related packages: We split up suitably formed sums
        of Feynman parameters at the cost of complex contour integrals
        \cite{Smirnov:1999gc,*Tausk:1999vh,Smirnov:2006ry,*Czakon:2005rk,
        *Smirnov:2009up}. Afterwards, the Feynman parameter integrals can
        usually be done in terms of Beta and Gamma functions, while the contour
        integrals give rise to infinite sums via the residue theorem.
  \item \emph{Almkvist-Zeilberger algorithm} \cite{Almkvist:1990,*Apagodu:2006,
        Ablinger:2013hcp}:
        If it is possible to confine the Mellin variable $N$ in the integrand of
        the Feynman parameter integrals to only one or a small number of places,
        it can be advantageous to employ the multi-variable Almkvist-Zeilberger
        algorithm, implemented in the package \progname{MultiIntegrate}
        \cite{Ablinger:2013hcp}. It allows to derive recurrence relations for
        the integrals which can subsequently be solved using the algorithms
        implemented in \progname{Sigma}.
  \item \emph{Differential equations and difference equations}
        \cite{Kotikov:1990kg,*Caffo:1998yd,*Caffo:1998du,*Gehrmann:1999as,
        Ablinger:2015tua}:
        Based on the integration-by-parts relations it is possible to derive
        coupled systems of differential equations for the master integrals.
        We translate these into coupled systems of difference equations and
        uncouple them using Z{\"u}rcher's algorithm \cite{Zuercher:1994}, which
        is implemented in \progname{OreSys} \cite{Gerhold:2002}. Once suitable
        initial conditions are available (e.g. from direct calculations using
        other methods), the difference equations can be solved using the package
        \progname{SolveCoupledSystem} \cite{Ablinger:2016pbw,*Ablinger:2016yjz}.
\end{itemize}
More details on the application of these techniques to the calculation of
massive OMEs can be found in \cite{Ablinger:2015tua,Ablinger:2010ty,Blumlein:2012vq,
Ablinger:2012qm,Ablinger:2014lka,Ablinger:2014uka,Ablinger:2014yaa}.
The results obtained so far for the OMEs can be expressed in terms of nested
sums. In particular harmonic sums \cite{Vermaseren:1998uu,*Blumlein:1998if,
*Blumlein:2000hw,*Blumlein:2003gb,*Blumlein:2009ta,*Blumlein:2009fz}, generalised
harmonic sums \cite{Moch:2001zr,Ablinger:2013cf}, cyclotomic sums
\cite{Ablinger:2011te} and binomially weighted sums \cite{Fleischer:1998nb,
*Davydychev:2003mv,*Weinzierl:2004bn,Ablinger:2014bra} appear both in the
intermediate steps and in the results. These structures are related to
corresponding iterated integrals \cite{Remiddi:1999ew,Moch:2001zr,
Ablinger:2013cf,Ablinger:2011te,Ablinger:2014bra} via an inverse Mellin
transformation. Finally, the results for the master integrals can be inserted
into the diagrams, yielding the unrenormalised expressions for the operator
matrix elements.

The renormalisation procedure for the massive OMEs at $\order{a_s^3}$ was worked
out in \cite{Bierenbaum:2009mv}. Since we calculate matrix elements of the local
light-cone operators, it comes at no surprise that their renormalisation involves
the anomalous dimensions of the operators. Using known results for the beta
function, mass anomalous dimensions and lower order OMEs, we can use the pole
terms of our 3-loop results to calculate the $N_F$-dependent part of the
anomalous dimensions.


\section{Results for structure functions}\label{sec:structure-functions}
Over the course of the recent years, a number of analytic results for the
operator matrix elements have been completed. In particular, the OMEs $\AqgQ{3}$
\cite{Ablinger:2010ty}, $\AgqQ{3}$ \cite{Ablinger:2014lka}, $\AqqQPS{3}$
\cite{Ablinger:2010ty}, $\AQqPS{3}$ \cite{Ablinger:2014nga}, $\AqqQNS{3}$
\cite{Ablinger:2014vwa} have been calculated. Moreover, the gluonic OME
$\AggQ{3}$ is known for even values of the Mellin variable $N$
\cite{DESY-15–112}. These results allow for a numerical illustration of their
impact on the heavy flavour Wilson coefficients of different structure
functions. For the structure function $F_2(x,Q^2)$ there are five different
Wilson coefficients (see Eq.~\eqref{eq:F2-wilson-coeff}), each of which requires
the knowledge of the 3-loop term of one OME. Since $\AQg{3}$ is not yet
fully known, the Wilson coefficient $H_{g,2}$ cannot be given yet at 3-loop order at
this point. Nevertheless, we can illustrate the impact of the remaining Wilson
coefficients, supplemented by the contribution of $H_{g,2}$ up to 2-loop order
for comparison.
\begin{figure}
  \centering
  \includegraphics{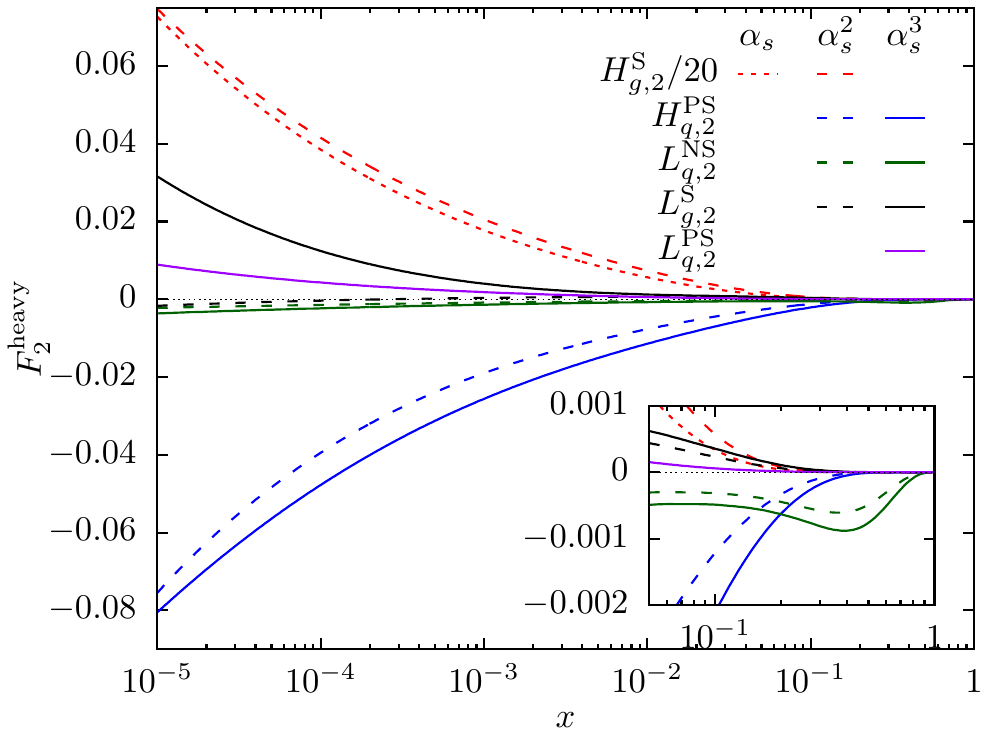}
  \caption{Illustration of the contributions to the structure function
           $F_2(x,Q^2)$ from the individual Wilson coefficients. The plot
           shows the values for $Q^2=100\,\mathrm{GeV}^2$ and the scale
           choice $\mu^2=Q^2$ with PDFs from \cite{Alekhin:2013nda}. The
           charm quark mass is $m_c=\SI{1.59}{\GeV^2}$ in the on-shell
           scheme \cite{Alekhin:2012vu}. The inset enlarges the region of
	   $0.05 \leq x \leq 1$.}
  \label{fig:F2-contributions}
\end{figure}
Figure \ref{fig:F2-contributions} shows the heavy flavour contributions to
$F_2(x,Q^2)$ for a fixed value of $Q^2=\SI{100}{\GeV^2}$.
The biggest contribution comes from $H_{g,2}$. Contributions to
this Wilson coefficient start at $\order{a_s}$ and are the only contribution at
that order. Due to this and the fact that it involves the gluon PDF, it is quite
large in the small-$x$ region and we have scaled down the curve by a factor
$20$.
The second largest contribution in the small-$x$ region is the pure-singlet
Wilson coefficient $H_{q,2}^\text{PS}$. It starts at $\mathcal{O}(a_s^2)$ and
is negative, except for very large values of $x$ (not visible in the plot).
The large-$x$ region is dominated by the non-singlet Wilson coefficient, which
also starts at $\mathcal{O}(a_s^2)$ and is negative throughout the whole
$x$-range. Here, the even moments of the non-singlet OME $\AqqQNS{3}$ enter.
Somewhat smaller contributions arise from the gluon- and quark-initiated singlet
Wilson coefficients $L_{g,2}^\text{S}$ and $L_{q,2}^\text{PS}$, which start at
$\mathcal{O}(a_s^2)$ and $\mathcal{O}(a_s^3)$, respectively.

\begin{figure}
  \centering
  \includegraphics[width=0.45\textwidth]{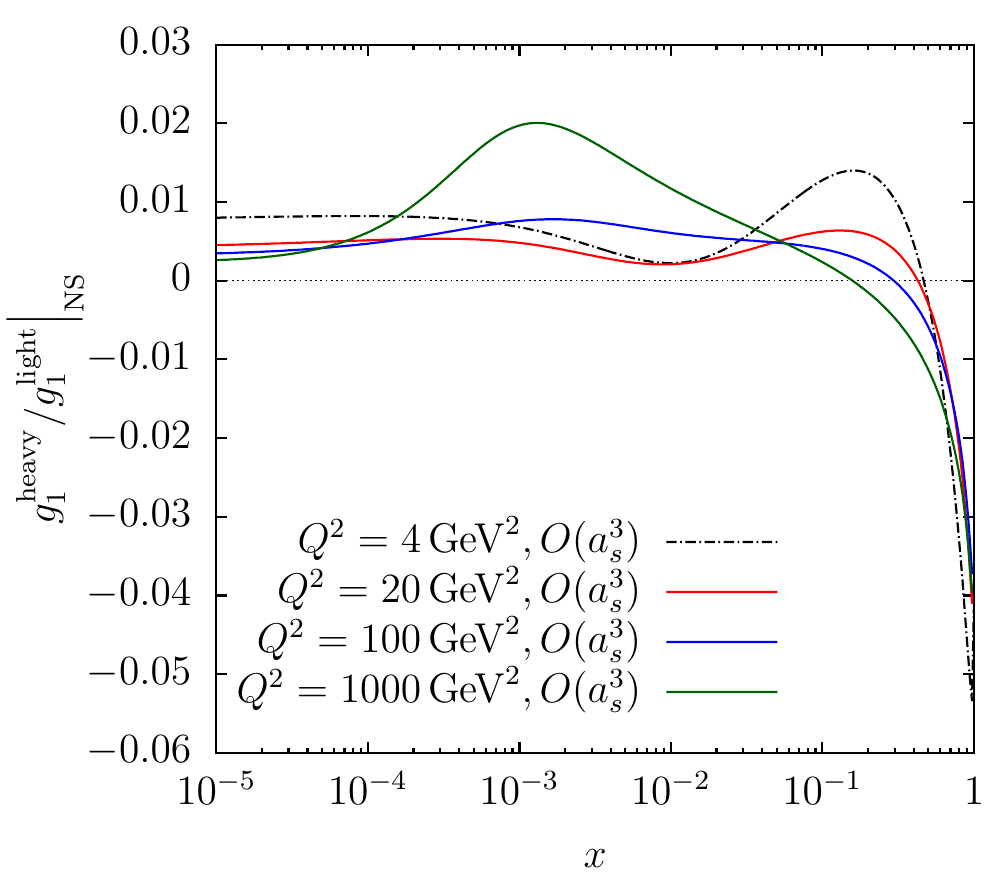}
  \hspace{2em}
  \includegraphics[width=0.45\textwidth]{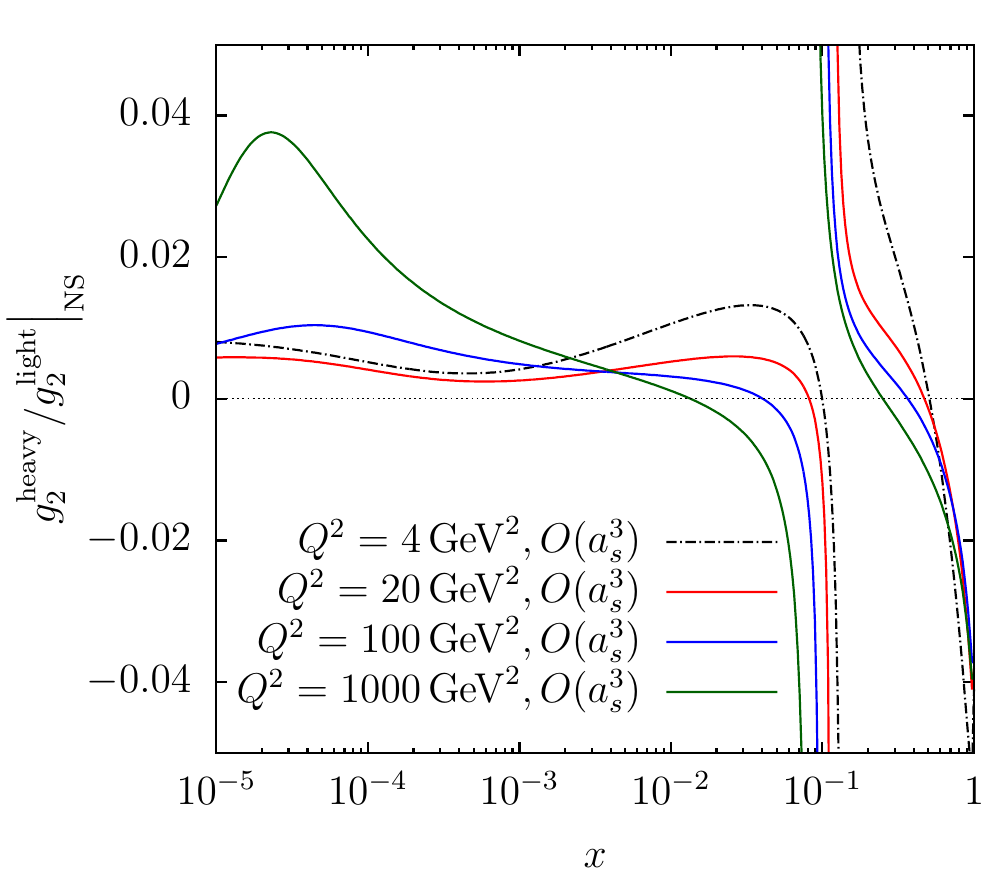}
  \caption{Ratio of the heavy over the light flavour contributions to
          the polarised structure functions $g_1(x,Q^2)$ (left panel)
          and $g_2(x,Q^2)$ (right panel) for different values of $Q^2$.
          Here, the PDFs from \cite{Blumlein:2010rn} were used; plots from
          \cite{Behring:2015zaa}.}
  \label{fig:g1g2}
\end{figure}
The non-singlet OME $\AqqQNS{3}$ was calculated also for odd moments in
\cite{Ablinger:2014vwa}. This allows for applications to other structure
functions besides $F_2(x,Q^2)$: In particular the non-singlet contribution to
the polarised structure function $g_1(x,Q^2)$ was explored in
\cite{Behring:2015zaa}. Figure \ref{fig:g1g2} illustrates the impact
of the charm quarks on this structure function. Compared to the contributions
from massless quarks and gluons, the charm quark constitutes around +1\% to 2\%
to -5\% of the non-singlet structure function. This is below the current experimental
accuracy but may be of interest at future high-luminosity colliders
\cite{Boer:2011fh,*Accardi:2012qut,*AbelleiraFernandez:2012cc}.

At the level of twist 2, the structure function $g_2(x,Q^2)$ is also related to
$g_1(x,Q^2)$ by the Wandzura-Wilczek relation \cite{Wandzura:1977qf},
\begin{align}
  g_2(x,Q^2) &= -g_1(x,Q^2) + \int_x^1 \frac{\mathrm{d} y}{y} g_1(y,Q^2)
  \label{eq:wandzura-wilczek}
  \,.
\end{align}
This allows us to show also the impact of the charm quarks there. As can be seen
in the right panel of Figure \ref{fig:g1g2}, the heavy quark contributions are
about $1\%$ to $4\%$ the size of the massless contribution. The pole in the plot
is due to a change of sign of $g_2^\text{light}$.

Another application of the non-singlet OME is the polarised Bjorken sum rule
\cite{Bjorken:1969mm}. It is defined as the difference of the first moments of
$g_1(x,Q^2)$ in electron-proton and electron-neutron scattering,
\begin{align}
  \int_0^1 \mathrm{d} x \left[g_1^{ep}(x,Q^2) - g_1^{en}(x,Q^2) \right]
    &= \frac{1}{6}\left|\frac{g_A}{g_V}\right| C_\text{pBj}(a_s)
  \label{eq:pol-bjorken}
  \,,
\end{align}
where $g_A$ and $g_V$ denote the axial-vector and vector decay constants. The
perturbative coefficient $C_\text{pBj}$ arises from first moment of the Wilson
coefficients. For the massless contributions it has been calculated up to
$\mathcal{O}(a_s^4)$ \cite{Baikov:2010je,Larin:2013yba,Baikov:2015tea}. Our
result for the non-singlet heavy flavour Wilson coefficient leads us to the
conclusion that in the limit $Q^2 \gg m^2$ the polarised Bjorken sum rule for
$N_F$ massless quarks and one massive quark is given completely by the massless
contributions for $N_F+1$ quarks: The massive non-singlet OME, which could
modify the sum rule compared to the completely massless case, has a vanishing
first moment due to fermion number conservation. However, away from the limit
$Q^2 \gg m^2$, the behaviour described above no longer holds and genuine heavy
flavour contributions to the polarised Bjorken sum rule arise, cf.
\cite{Blumlein:2016xcy}.

\begin{figure}
  \centering
  \includegraphics[width=0.45\textwidth]{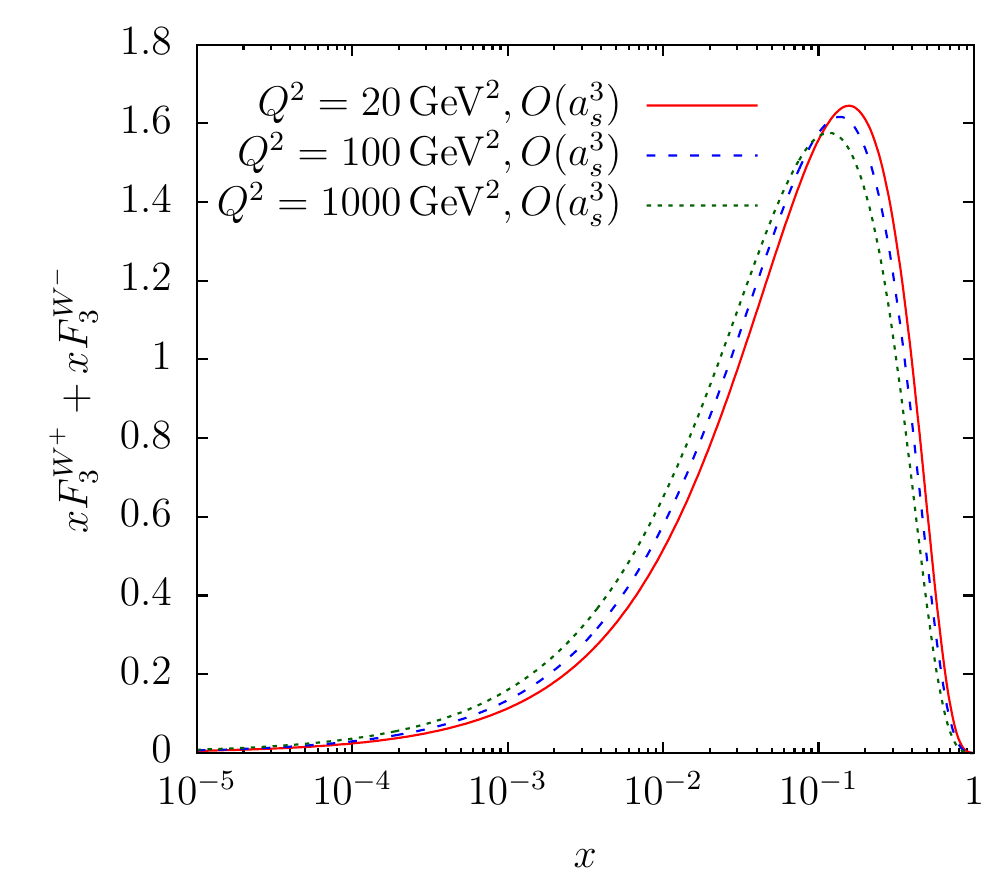}
  \hspace{2em}
  \includegraphics[width=0.45\textwidth]{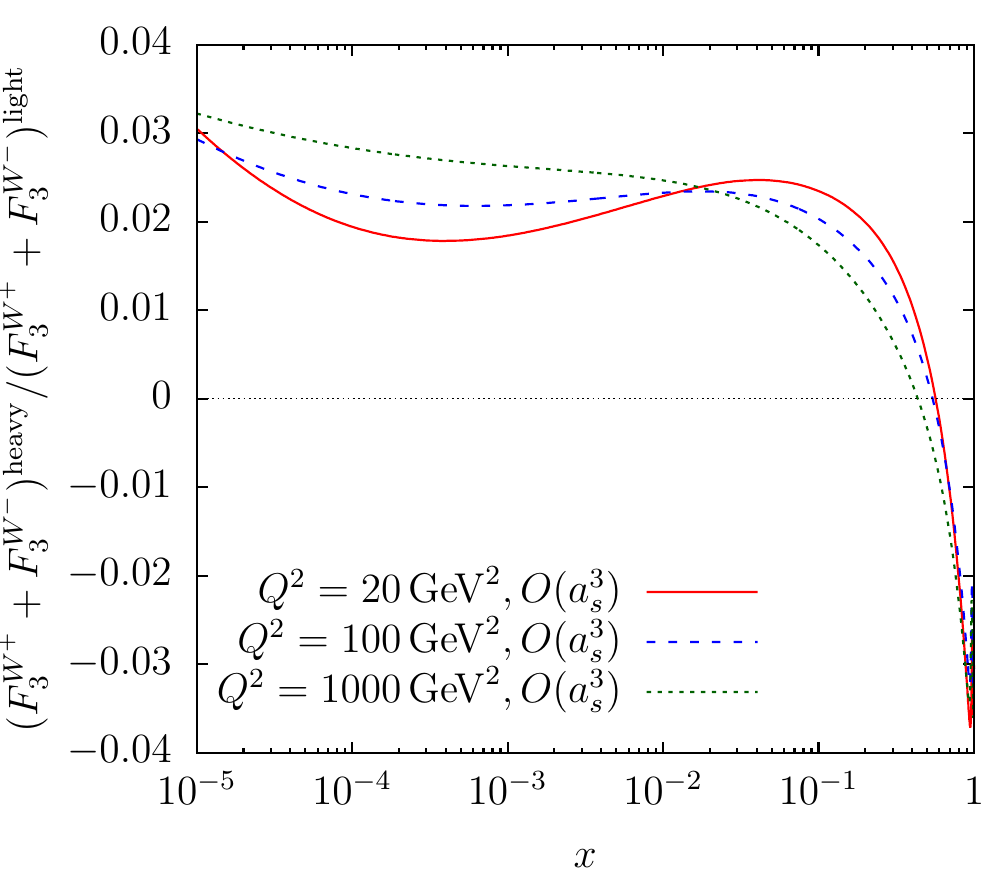}
  \caption{\emph{Left panel}: Illustration of the structure function combination
           $x F_3^{W^+ - W^-}(x,Q^2)$ including the contributions from a massive
           charm quark with a mass of $m_c=\SI{1.59}{\GeV^2}$ in the on-shell
           scheme \cite{Alekhin:2012vu}.
           \emph{Right panel}: Ratio of the heavy quark contributions over the
           massless part of the same structure function;
           plots from \cite{Behring:2015roa}.}
  \label{fig:xF3}
\end{figure}
Moreover, the same non-singlet OME enters also the charged-current structure
function combination
\begin{align}
  x F_3^{W^+ - W^-}(x,Q^2) &= x F_3^{W^+}(x,Q^2) + x F_3^{W^-}(x,Q^2)
  \label{eq:xF3WmW}
  \,.
\end{align}
For an explanation of the notation we refer to \cite{Blumlein:2011zu,
*Blumlein:2014fqa,Behring:2015roa}. This structure function receives
contributions from two non-singlet Wilson coefficients:
On the one hand, there is $L_{q,3}^\text{NS}$, which describes reactions in
which the $W$ boson couples to a light quark and mediates a flavour transition
between light quark species. This is analogous to the case of photon exchange,
except for the flavour change.
On the other hand, there is $H_{q,3}^\text{NS}$, which describes flavour
excitation reactions (e.g. $s \to c$). Here, the $W$ boson couples to a heavy
quark. This part has no analogy in the photon-mediated case.
The impact of both Wilson coefficients on $x F_3^{W^+ - W^-}(x,Q^2)$ is
illustrated in Figure \ref{fig:xF3}. The size of the heavy quark contribution
is again of the order of about $3\%$.
Also in the charged current sector, there is a sum rule arising from the first
moment of the structure functions: The Gross-Llewellyn-Smith sum rule
\cite{Gross:1969jf} is given by
\begin{align}
  \int_0^1 \mathrm{d} x \left[F_3^{\bar{\nu}p}(x,Q^2)+F_3^{\nu p}(x,Q^2)\right]
    &= 6 \, C_\text{GLS}(a_s)
  \label{eq:gls-sum-rule}
  \,.
\end{align}
The massless QCD corrections to $C_\text{GLS}$ are again known to
$\mathcal{O}(a_s^4)$ \cite{Baikov:2010je,Baikov:2010iw,Baikov:2012zn} and the
situation in the heavy quark sector is similar to the polarised case: Due to the
vanishing first moment of $A_{qq,Q}^\text{NS}$, the influence of the heavy quark
in the asymptotic region $Q^2 \gg m^2$ reduces to incrementing the number of
massless flavours ($N_F \to N_F+1$). For the power corrections in $m^2/Q^2$ up
to 2-loop order, we refer to \cite{Blumlein:2016xcy}.


\section{Operator matrix element for transversity}\label{sec:transversity}
The non-singlet results discussed so far involve the flavour non-singlet
vector or axial-vector operators. A very similar calculation can be performed
for the non-singlet tensor operator
\begin{align}
  O_{q,r}^{\text{TR,NS},\mu\mu_1\dots\mu_N}(z) &=
    \mathrm{i}^{N-1} S\left[\bar{\psi}(z) \sigma^{\mu\mu_1} D^{\mu_2}
      \dots D^{\mu_N} \frac{\lambda_r}{2} \psi(z)\right] - \text{trace terms}
  \label{eq:operator-transversity}
  \,,
\end{align}
where $D^\mu$ is the covariant derivative, $\psi$ is the quark field operator,
$\lambda_r$ denotes the Gell-Mann matrices of $\mathrm{SU}(3)_\text{flavour}$,
$\sigma^{\mu\nu} = \tfrac{\mathrm{i}}{2} (\gamma^\mu\gamma^\nu -
\gamma^\nu\gamma^\mu)$ and $S[\dots]$ denotes the symmetrisation of the Lorentz
indices.

The operator in Eq. \eqref{eq:operator-transversity} is related to the
transversity distribution $\Delta_T f_k$ which appears in the transversity
structure function $h_1(x,Q^2)$ \cite{Barone:2001sp}. It can be measured in
semi-inclusive deep-inelastic scattering \cite{Ralston:1979ys,*Jaffe:1991kp,
*Jaffe:1991ra,*Cortes:1991ja} and polarised Drell-Yan processes
\cite{Artru:1989zv,*Collins:1992kk,*Jaffe:1993xb,*Tangerman:1994bb,
*Boer:1997nt}.
The massive OME of this operator is denoted by $\AqqQTR{3}$ and involves the
same diagrams as the vector non-singlet operator $\AqqQNS{3}$.
Of course, different Feynman rules and a different
projector have to be used, but the required master integrals turn out to be the
same. Our calculation allows to extract the $N_F$-dependent parts of the
anomalous dimensions of the transversity operator up to 3-loop order.
The result for the anomalous dimensions and the complete OME $\AqqQTR{3}$ are
given in \cite{Ablinger:2014vwa}. In principle the heavy flavour Wilson
coefficients could be constructed in a similar fashion as for the other
structure functions, but at this point the light flavour Wilson coefficients
have not been calculated to a sufficient order yet.


\section{Conclusions}\label{sec:conclusions}
Up to now, seven out of eight 3-loop massive operator matrix elements have
been calculated analytically for general values of the Mellin variable $N$.
Here, we discussed both the framework of calculation as well as a selected
number of applications.
The analytic calculation of the required 3-loop master integrals with
additional local operator insertions led to the development and improvement of
computer-algebraic and mathematical methods and tools. The packages
\progname{Sigma}, \progname{HarmonicSums}, \progname{EvaluateMultiSums},
\progname{SumProduction} and \progname{SolveCoupledSystem} both enabled the
calculation and have benefited greatly from the challenges posed by the tasks
arising from the project.

The largest contribution to the heavy flavour corrections to the structure
function $F_2(x,Q^2)$ is expected to come from the Wilson coefficient
$H_{g,2}^\text{S}$, which is not yet known due to the missing OME $\AQg{3}$.
Nonetheless, a first impression of the impact of the heavy flavour corrections
can be seen from our illustrations in Section \ref{sec:structure-functions}.
We also discussed the influence of the massive OME $\AqqQNS{3}$ on the
non-singlet polarised and charged-current structure functions $g_1(x,Q^2)$ and
$x F_3(x,Q^2)$, which is of the order of a few percent in both cases. Due to
the vanishing first moment of the non-singlet OME, the polarised Bjorken and
Gross-Llewellyn-Smith sum rules are only modified by shifting $N_F$ to $N_F+1$
if compared to the purely massless case with $N_F$ quark flavours in the region $Q^2 \gg m^2$.
Finally, we also calculated the massive OME for the non-singlet transversity
operator, which enabled us to obtain the $N_F$-dependent parts of the 3-loop
anomalous dimension of this operator.

\bibliographystyle{JHEPM}
\bibliography{lit}

\ifx\mcitethebibliography\mciteundefinedmacro
\PackageError{JHEPM.bst}{mciteplus.sty has not been loaded}
{This bibstyle requires the use of the mciteplus package.}\fi
\providecommand{\href}[2]{#2}\begingroup\raggedright\begin{mcitethebibliography}{100}

\bibitem{Bethke:2011tr}
S.~Bethke et~al., \emph{{Workshop on Precision Measurements of $\alpha_s$}},
  \href{https://arxiv.org/abs/1110.0016}{{\tt 1110.0016}}\relax
\mciteBstWouldAddEndPuncttrue
\mciteSetBstMidEndSepPunct{\mcitedefaultmidpunct}
{\mcitedefaultendpunct}{\mcitedefaultseppunct}\relax
\EndOfBibitem
\bibitem{Moch:2014tta}
S.~Moch et~al., \emph{{High precision fundamental constants at the TeV scale}},
   \href{https://arxiv.org/abs/1405.4781}{{\tt 1405.4781}}\relax
\mciteBstWouldAddEndPuncttrue
\mciteSetBstMidEndSepPunct{\mcitedefaultmidpunct}
{\mcitedefaultendpunct}{\mcitedefaultseppunct}\relax
\EndOfBibitem
\bibitem{Alekhin:2016evh}
S.~Alekhin, J.~B{\"u}mlein and S.~O. Moch, \emph{{$\alpha_s$ from global fits
  of parton distribution functions}},
  \href{http://dx.doi.org/10.1142/S0217732316300238}{\emph{Mod. Phys. Lett.}
  {\bf A31} (2016) 1630023}\relax
\mciteBstWouldAddEndPuncttrue
\mciteSetBstMidEndSepPunct{\mcitedefaultmidpunct}
{\mcitedefaultendpunct}{\mcitedefaultseppunct}\relax
\EndOfBibitem
\bibitem{Alekhin:2013nda}
S.~Alekhin, J.~Bl{\"u}mlein and S.~Moch, \emph{{The ABM parton distributions
  tuned to LHC data}},
  \href{http://dx.doi.org/10.1103/PhysRevD.89.054028}{\emph{Phys. Rev.} {\bf
  D89} (2014) 054028}, [\href{https://arxiv.org/abs/1310.3059}{{\tt
  1310.3059}}]\relax
\mciteBstWouldAddEndPuncttrue
\mciteSetBstMidEndSepPunct{\mcitedefaultmidpunct}
{\mcitedefaultendpunct}{\mcitedefaultseppunct}\relax
\EndOfBibitem
\bibitem{Jimenez-Delgado:2014twa}
P.~Jimenez-Delgado and E.~Reya, \emph{{Delineating parton distributions and the
  strong coupling}},
  \href{http://dx.doi.org/10.1103/PhysRevD.89.074049}{\emph{Phys. Rev.} {\bf
  D89} (2014) 074049}, [\href{https://arxiv.org/abs/1403.1852}{{\tt
  1403.1852}}]\relax
\mciteBstWouldAddEndPuncttrue
\mciteSetBstMidEndSepPunct{\mcitedefaultmidpunct}
{\mcitedefaultendpunct}{\mcitedefaultseppunct}\relax
\EndOfBibitem
\bibitem{Dulat:2015mca}
S.~Dulat, T.-J. Hou, J.~Gao, M.~Guzzi, J.~Huston, P.~Nadolsky et~al.,
  \emph{{New parton distribution functions from a global analysis of quantum
  chromodynamics}},
  \href{http://dx.doi.org/10.1103/PhysRevD.93.033006}{\emph{Phys. Rev.} {\bf
  D93} (2016) 033006}, [\href{https://arxiv.org/abs/1506.07443}{{\tt
  1506.07443}}]\relax
\mciteBstWouldAddEndPuncttrue
\mciteSetBstMidEndSepPunct{\mcitedefaultmidpunct}
{\mcitedefaultendpunct}{\mcitedefaultseppunct}\relax
\EndOfBibitem
\bibitem{Harland-Lang:2014zoa}
L.~A. Harland-Lang, A.~D. Martin, P.~Motylinski and R.~S. Thorne, \emph{{Parton
  distributions in the LHC era: MMHT 2014 PDFs}},
  \href{http://dx.doi.org/10.1140/epjc/s10052-015-3397-6}{\emph{Eur. Phys. J.}
  {\bf C75} (2015) 204}, [\href{https://arxiv.org/abs/1412.3989}{{\tt
  1412.3989}}]\relax
\mciteBstWouldAddEndPuncttrue
\mciteSetBstMidEndSepPunct{\mcitedefaultmidpunct}
{\mcitedefaultendpunct}{\mcitedefaultseppunct}\relax
\EndOfBibitem
\bibitem{Ball:2014uwa}
{\scshape NNPDF} collaboration, R.~D. Ball et~al., \emph{{Parton distributions
  for the LHC Run II}},
  \href{http://dx.doi.org/10.1007/JHEP04(2015)040}{\emph{JHEP} {\bf 04} (2015)
  040}, [\href{https://arxiv.org/abs/1410.8849}{{\tt 1410.8849}}]\relax
\mciteBstWouldAddEndPuncttrue
\mciteSetBstMidEndSepPunct{\mcitedefaultmidpunct}
{\mcitedefaultendpunct}{\mcitedefaultseppunct}\relax
\EndOfBibitem
\bibitem{Abramowicz:2015mha}
{\scshape ZEUS, H1} collaboration, H.~Abramowicz et~al., \emph{{Combination of
  measurements of inclusive deep inelastic ${e^{\pm }p}$ scattering cross
  sections and QCD analysis of HERA data}},
  \href{http://dx.doi.org/10.1140/epjc/s10052-015-3710-4}{\emph{Eur. Phys. J.}
  {\bf C75} (2015) 580}, [\href{https://arxiv.org/abs/1506.06042}{{\tt
  1506.06042}}]\relax
\mciteBstWouldAddEndPuncttrue
\mciteSetBstMidEndSepPunct{\mcitedefaultmidpunct}
{\mcitedefaultendpunct}{\mcitedefaultseppunct}\relax
\EndOfBibitem
\bibitem{Accardi:2016ndt}
A.~Accardi et~al., \emph{{A Critical Appraisal and Evaluation of Modern PDFs}},
  \href{http://dx.doi.org/10.1140/epjc/s10052-016-4285-4}{\emph{Eur. Phys. J.}
  {\bf C76} (2016) 471}, [\href{https://arxiv.org/abs/1603.08906}{{\tt
  1603.08906}}]\relax
\mciteBstWouldAddEndPuncttrue
\mciteSetBstMidEndSepPunct{\mcitedefaultmidpunct}
{\mcitedefaultendpunct}{\mcitedefaultseppunct}\relax
\EndOfBibitem
\bibitem{Alekhin:2012vu}
S.~Alekhin, J.~Bl{\"u}mlein, K.~Daum, K.~Lipka and S.~Moch, \emph{{Precise
  charm-quark mass from deep-inelastic scattering}},
  \href{http://dx.doi.org/10.1016/j.physletb.2013.02.010}{\emph{Phys. Lett.}
  {\bf B720} (2013) 172--176}, [\href{https://arxiv.org/abs/1212.2355}{{\tt
  1212.2355}}]\relax
\mciteBstWouldAddEndPuncttrue
\mciteSetBstMidEndSepPunct{\mcitedefaultmidpunct}
{\mcitedefaultendpunct}{\mcitedefaultseppunct}\relax
\EndOfBibitem
\bibitem{Alekhin:2016uxn}
S.~Alekhin, J.~Bl{\"u}mlein, S.-O. Moch and R.~Pla{\v c}akyt{\.e}, \emph{{The
  new ABMP16 PDF}}, {\emph{PoS} {\bf DIS2016} (2016) 016},
  [\href{https://arxiv.org/abs/1609.03327}{{\tt 1609.03327}}]\relax
\mciteBstWouldAddEndPuncttrue
\mciteSetBstMidEndSepPunct{\mcitedefaultmidpunct}
{\mcitedefaultendpunct}{\mcitedefaultseppunct}\relax
\EndOfBibitem
\bibitem{Buza:1995ie}
M.~Buza, Y.~Matiounine, J.~Smith, R.~Migneron and W.~L. van Neerven,
  \emph{{Heavy quark coefficient functions at asymptotic values $Q^2 \gg
  m^2$}}, \href{http://dx.doi.org/10.1016/0550-3213(96)00228-3}{\emph{Nucl.
  Phys.} {\bf B472} (1996) 611--658},
  [\href{https://arxiv.org/abs/hep-ph/9601302}{{\tt hep-ph/9601302}}]\relax
\mciteBstWouldAddEndPuncttrue
\mciteSetBstMidEndSepPunct{\mcitedefaultmidpunct}
{\mcitedefaultendpunct}{\mcitedefaultseppunct}\relax
\EndOfBibitem
\bibitem{Buras:1979yt}
A.~J. Buras, \emph{{Asymptotic Freedom in Deep Inelastic Processes in the
  Leading Order and Beyond}},
  \href{http://dx.doi.org/10.1103/RevModPhys.52.199}{\emph{Rev. Mod. Phys.}
  {\bf 52} (1980) 199}\relax
\mciteBstWouldAddEndPuncttrue
\mciteSetBstMidEndSepPunct{\mcitedefaultmidpunct}
{\mcitedefaultendpunct}{\mcitedefaultseppunct}\relax
\EndOfBibitem
\bibitem{Reya:1979zk}
E.~Reya, \emph{{Perturbative Quantum Chromodynamics}},
  \href{http://dx.doi.org/10.1016/0370-1573(81)90036-3}{\emph{Phys. Rept.} {\bf
  69} (1981) 195}\relax
\mciteBstWouldAddEndPuncttrue
\mciteSetBstMidEndSepPunct{\mcitedefaultmidpunct}
{\mcitedefaultendpunct}{\mcitedefaultseppunct}\relax
\EndOfBibitem
\bibitem{Blumlein:2012bf}
J.~Bl{\"u}mlein, \emph{{The Theory of Deeply Inelastic Scattering}},
  \href{http://dx.doi.org/10.1016/j.ppnp.2012.09.006}{\emph{Prog. Part. Nucl.
  Phys.} {\bf 69} (2013) 28--84}, [\href{https://arxiv.org/abs/1208.6087}{{\tt
  1208.6087}}]\relax
\mciteBstWouldAddEndPuncttrue
\mciteSetBstMidEndSepPunct{\mcitedefaultmidpunct}
{\mcitedefaultendpunct}{\mcitedefaultseppunct}\relax
\EndOfBibitem
\bibitem{Ablinger:2011pb}
J.~Ablinger, J.~Bl{\"u}mlein, S.~Klein, C.~Schneider and F.~Wi{\ss}brock,
  \emph{{3-Loop Heavy Flavor Corrections to DIS with two Massive Fermion
  Lines}},  in \emph{{19th International Workshop on Deep-Inelastic Scattering
  and Related Subjects (DIS 2011) Newport News, Virginia, April 11-15, 2011}},
  2011.
\newblock \href{https://arxiv.org/abs/1106.5937}{{\tt 1106.5937}}\relax
\mciteBstWouldAddEndPuncttrue
\mciteSetBstMidEndSepPunct{\mcitedefaultmidpunct}
{\mcitedefaultendpunct}{\mcitedefaultseppunct}\relax
\EndOfBibitem
\bibitem{Ablinger:2012qj}
J.~Ablinger, J.~Bl{\"u}mlein, A.~Hasselhuhn, S.~Klein, C.~Schneider and
  F.~Wi{\ss}brock, \emph{{New Heavy Flavor Contributions to the DIS Structure
  Function $F_2(x,Q^2)$ at $\mathcal{O}(\alpha_s^3)$}}, {\emph{PoS} {\bf
  RADCOR2011} (2011) 031}, [\href{https://arxiv.org/abs/1202.2700}{{\tt
  1202.2700}}]\relax
\mciteBstWouldAddEndPuncttrue
\mciteSetBstMidEndSepPunct{\mcitedefaultmidpunct}
{\mcitedefaultendpunct}{\mcitedefaultseppunct}\relax
\EndOfBibitem
\bibitem{DESY-14-019}
J.~Ablinger, J.~Bl{\"u}mlein, A.~De~Freitas, A.~Hasselhuhn, C.~Schneider and
  F.~Wi{\ss}brock. DESY-14-019\relax
\mciteBstWouldAddEndPuncttrue
\mciteSetBstMidEndSepPunct{\mcitedefaultmidpunct}
{\mcitedefaultendpunct}{\mcitedefaultseppunct}\relax
\EndOfBibitem
\bibitem{Bierenbaum:2009mv}
I.~Bierenbaum, J.~Bl{\"u}mlein and S.~Klein, \emph{{Mellin Moments of the
  $O(\alpha_s^3)$ Heavy Flavor Contributions to unpolarized Deep-Inelastic
  Scattering at $Q^2 \gg m^2$ and Anomalous Dimensions}},
  \href{http://dx.doi.org/10.1016/j.nuclphysb.2009.06.005}{\emph{Nucl. Phys.}
  {\bf B820} (2009) 417--482}, [\href{https://arxiv.org/abs/0904.3563}{{\tt
  0904.3563}}]\relax
\mciteBstWouldAddEndPuncttrue
\mciteSetBstMidEndSepPunct{\mcitedefaultmidpunct}
{\mcitedefaultendpunct}{\mcitedefaultseppunct}\relax
\EndOfBibitem
\bibitem{Behring:2014eya}
A.~Behring, I.~Bierenbaum, J.~Bl{\"u}mlein, A.~De~Freitas, S.~Klein and
  F.~Wi{\ss}brock, \emph{{The logarithmic contributions to the $O(\alpha^3_s)$
  asymptotic massive Wilson coefficients and operator matrix elements in deeply
  inelastic scattering}},
  \href{http://dx.doi.org/10.1140/epjc/s10052-014-3033-x}{\emph{Eur. Phys. J.}
  {\bf C74} (2014) 3033}, [\href{https://arxiv.org/abs/1403.6356}{{\tt
  1403.6356}}]\relax
\mciteBstWouldAddEndPuncttrue
\mciteSetBstMidEndSepPunct{\mcitedefaultmidpunct}
{\mcitedefaultendpunct}{\mcitedefaultseppunct}\relax
\EndOfBibitem
\bibitem{Vermaseren:2005qc}
J.~A.~M. Vermaseren, A.~Vogt and S.~Moch, \emph{{The third-order QCD
  corrections to deep-inelastic scattering by photon exchange}},
  \href{http://dx.doi.org/10.1016/j.nuclphysb.2005.06.020}{\emph{Nucl. Phys.}
  {\bf B724} (2005) 3--182}, [\href{https://arxiv.org/abs/hep-ph/0504242}{{\tt
  hep-ph/0504242}}]\relax
\mciteBstWouldAddEndPuncttrue
\mciteSetBstMidEndSepPunct{\mcitedefaultmidpunct}
{\mcitedefaultendpunct}{\mcitedefaultseppunct}\relax
\EndOfBibitem
\bibitem{Laenen:1992zk}
E.~Laenen, S.~Riemersma, J.~Smith and W.~L. van Neerven, \emph{{Complete
  $O(\alpha_s)$ corrections to heavy flavor structure functions in
  electroproduction}},
  \href{http://dx.doi.org/10.1016/0550-3213(93)90201-Y}{\emph{Nucl. Phys.} {\bf
  B392} (1993) 162--228}\relax
\mciteBstWouldAddEndPuncttrue
\mciteSetBstMidEndSepPunct{\mcitedefaultmidpunct}
{\mcitedefaultendpunct}{\mcitedefaultseppunct}\relax
\EndOfBibitem
\bibitem{Laenen:1992xs}
E.~Laenen, S.~Riemersma, J.~Smith and W.~L. van Neerven, \emph{{$O(\alpha_s)$
  corrections to heavy flavor inclusive distributions in electroproduction}},
  \href{http://dx.doi.org/10.1016/0550-3213(93)90202-Z}{\emph{Nucl. Phys.} {\bf
  B392} (1993) 229--250}\relax
\mciteBstWouldAddEndPuncttrue
\mciteSetBstMidEndSepPunct{\mcitedefaultmidpunct}
{\mcitedefaultendpunct}{\mcitedefaultseppunct}\relax
\EndOfBibitem
\bibitem{Riemersma:1994hv}
S.~Riemersma, J.~Smith and W.~L. van Neerven, \emph{{Rates for inclusive deep
  inelastic electroproduction of charm quarks at HERA}},
  \href{http://dx.doi.org/10.1016/0370-2693(95)00036-K}{\emph{Phys. Lett.} {\bf
  B347} (1995) 143--151}, [\href{https://arxiv.org/abs/hep-ph/9411431}{{\tt
  hep-ph/9411431}}]\relax
\mciteBstWouldAddEndPuncttrue
\mciteSetBstMidEndSepPunct{\mcitedefaultmidpunct}
{\mcitedefaultendpunct}{\mcitedefaultseppunct}\relax
\EndOfBibitem
\bibitem{Alekhin:2003ev}
S.~I. Alekhin and J.~Bl{\"u}mlein, \emph{{Mellin representation for the heavy
  flavor contributions to deep inelastic structure functions}},
  \href{http://dx.doi.org/10.1016/j.physletb.2004.05.042}{\emph{Phys. Lett.}
  {\bf B594} (2004) 299--307},
  [\href{https://arxiv.org/abs/hep-ph/0404034}{{\tt hep-ph/0404034}}]\relax
\mciteBstWouldAddEndPuncttrue
\mciteSetBstMidEndSepPunct{\mcitedefaultmidpunct}
{\mcitedefaultendpunct}{\mcitedefaultseppunct}\relax
\EndOfBibitem
\bibitem{Buza:1996wv}
M.~Buza, Y.~Matiounine, J.~Smith and W.~L. van Neerven, \emph{{Charm
  electroproduction viewed in the variable flavor number scheme versus fixed
  order perturbation theory}},
  \href{http://dx.doi.org/10.1007/BF01245820}{\emph{Eur. Phys. J.} {\bf C1}
  (1998) 301--320}, [\href{https://arxiv.org/abs/hep-ph/9612398}{{\tt
  hep-ph/9612398}}]\relax
\mciteBstWouldAddEndPuncttrue
\mciteSetBstMidEndSepPunct{\mcitedefaultmidpunct}
{\mcitedefaultendpunct}{\mcitedefaultseppunct}\relax
\EndOfBibitem
\bibitem{Buza:1996xr}
M.~Buza, Y.~Matiounine, J.~Smith and W.~L. van Neerven, \emph{{$O(\alpha_s^2)$
  corrections to polarized heavy flavor production at $Q^2 \gg m^2$}},
  \href{http://dx.doi.org/10.1016/S0550-3213(96)00606-2}{\emph{Nucl. Phys.}
  {\bf B485} (1997) 420--456},
  [\href{https://arxiv.org/abs/hep-ph/9608342}{{\tt hep-ph/9608342}}]\relax
\mciteBstWouldAddEndPuncttrue
\mciteSetBstMidEndSepPunct{\mcitedefaultmidpunct}
{\mcitedefaultendpunct}{\mcitedefaultseppunct}\relax
\EndOfBibitem
\bibitem{Bierenbaum:2007qe}
I.~Bierenbaum, J.~Bl{\"u}mlein and S.~Klein, \emph{{Two-Loop Massive Operator
  Matrix Elements and Unpolarized Heavy Flavor Production at Asymptotic Values
  $Q^2 \gg m^2$}},
  \href{http://dx.doi.org/10.1016/j.nuclphysb.2007.04.030}{\emph{Nucl. Phys.}
  {\bf B780} (2007) 40--75}, [\href{https://arxiv.org/abs/hep-ph/0703285}{{\tt
  hep-ph/0703285}}]\relax
\mciteBstWouldAddEndPuncttrue
\mciteSetBstMidEndSepPunct{\mcitedefaultmidpunct}
{\mcitedefaultendpunct}{\mcitedefaultseppunct}\relax
\EndOfBibitem
\bibitem{Bierenbaum:2008yu}
I.~Bierenbaum, J.~Bl{\"u}mlein, S.~Klein and C.~Schneider, \emph{{Two-Loop
  Massive Operator Matrix Elements for Unpolarized Heavy Flavor Production to
  $O(\varepsilon)$}},
  \href{http://dx.doi.org/10.1016/j.nuclphysb.2008.05.016}{\emph{Nucl. Phys.}
  {\bf B803} (2008) 1--41}, [\href{https://arxiv.org/abs/0803.0273}{{\tt
  0803.0273}}]\relax
\mciteBstWouldAddEndPuncttrue
\mciteSetBstMidEndSepPunct{\mcitedefaultmidpunct}
{\mcitedefaultendpunct}{\mcitedefaultseppunct}\relax
\EndOfBibitem
\bibitem{Bierenbaum:2009zt}
I.~Bierenbaum, J.~Bl{\"u}mlein and S.~Klein, \emph{{The Gluonic Operator Matrix
  Elements at $O(\alpha_s^2)$ for DIS Heavy Flavor Production}},
  \href{http://dx.doi.org/10.1016/j.physletb.2009.01.057}{\emph{Phys. Lett.}
  {\bf B672} (2009) 401--406}, [\href{https://arxiv.org/abs/0901.0669}{{\tt
  0901.0669}}]\relax
\mciteBstWouldAddEndPuncttrue
\mciteSetBstMidEndSepPunct{\mcitedefaultmidpunct}
{\mcitedefaultendpunct}{\mcitedefaultseppunct}\relax
\EndOfBibitem
\bibitem{tHooft:1972fi}
G.~'t~Hooft and M.~J.~G. Veltman, \emph{{Regularization and Renormalization of
  Gauge Fields}},
  \href{http://dx.doi.org/10.1016/0550-3213(72)90279-9}{\emph{Nucl. Phys.} {\bf
  B44} (1972) 189--213}\relax
\mciteBstWouldAddEndPuncttrue
\mciteSetBstMidEndSepPunct{\mcitedefaultmidpunct}
{\mcitedefaultendpunct}{\mcitedefaultseppunct}\relax
\EndOfBibitem
\bibitem{Ashmore:1972uj}
J.~F. Ashmore, \emph{{A Method of Gauge Invariant Regularization}},
  \href{http://dx.doi.org/10.1007/BF02824407}{\emph{Lett. Nuovo Cim.} {\bf 4}
  (1972) 289--290}\relax
\mciteBstWouldAddEndPuncttrue
\mciteSetBstMidEndSepPunct{\mcitedefaultmidpunct}
{\mcitedefaultendpunct}{\mcitedefaultseppunct}\relax
\EndOfBibitem
\bibitem{Cicuta:1972jf}
G.~M. Cicuta and E.~Montaldi, \emph{{Analytic renormalization via continuous
  space dimension}}, \href{http://dx.doi.org/10.1007/BF02756527}{\emph{Lett.
  Nuovo Cim.} {\bf 4} (1972) 329--332}\relax
\mciteBstWouldAddEndPuncttrue
\mciteSetBstMidEndSepPunct{\mcitedefaultmidpunct}
{\mcitedefaultendpunct}{\mcitedefaultseppunct}\relax
\EndOfBibitem
\bibitem{Bollini:1972ui}
C.~G. Bollini and J.~J. Giambiagi, \emph{{Dimensional Renormalization: The
  Number of Dimensions as a Regularizing Parameter}},
  \href{http://dx.doi.org/10.1007/BF02895558}{\emph{Nuovo Cim.} {\bf B12}
  (1972) 20--26}\relax
\mciteBstWouldAddEndPuncttrue
\mciteSetBstMidEndSepPunct{\mcitedefaultmidpunct}
{\mcitedefaultendpunct}{\mcitedefaultseppunct}\relax
\EndOfBibitem
\bibitem{Nogueira:1991ex}
P.~Nogueira, \emph{{Automatic Feynman graph generation}},
  \href{http://dx.doi.org/10.1006/jcph.1993.1074}{\emph{J. Comput. Phys.} {\bf
  105} (1993) 279--289}\relax
\mciteBstWouldAddEndPuncttrue
\mciteSetBstMidEndSepPunct{\mcitedefaultmidpunct}
{\mcitedefaultendpunct}{\mcitedefaultseppunct}\relax
\EndOfBibitem
\bibitem{Vermaseren:2000nd}
J.~A.~M. Vermaseren, \emph{{New features of FORM}},
  \href{https://arxiv.org/abs/math-ph/0010025}{{\tt math-ph/0010025}}\relax
\mciteBstWouldAddEndPuncttrue
\mciteSetBstMidEndSepPunct{\mcitedefaultmidpunct}
{\mcitedefaultendpunct}{\mcitedefaultseppunct}\relax
\EndOfBibitem
\bibitem{Tentyukov:2007mu}
M.~Tentyukov and J.~A.~M. Vermaseren, \emph{{The Multithreaded version of
  FORM}}, \href{http://dx.doi.org/10.1016/j.cpc.2010.04.009}{\emph{Comput.
  Phys. Commun.} {\bf 181} (2010) 1419--1427},
  [\href{https://arxiv.org/abs/hep-ph/0702279}{{\tt hep-ph/0702279}}]\relax
\mciteBstWouldAddEndPuncttrue
\mciteSetBstMidEndSepPunct{\mcitedefaultmidpunct}
{\mcitedefaultendpunct}{\mcitedefaultseppunct}\relax
\EndOfBibitem
\bibitem{Kuipers:2012rf}
J.~Kuipers, T.~Ueda, J.~A.~M. Vermaseren and J.~Vollinga, \emph{{FORM version
  4.0}}, \href{http://dx.doi.org/10.1016/j.cpc.2012.12.028}{\emph{Comput. Phys.
  Commun.} {\bf 184} (2013) 1453--1467},
  [\href{https://arxiv.org/abs/1203.6543}{{\tt 1203.6543}}]\relax
\mciteBstWouldAddEndPuncttrue
\mciteSetBstMidEndSepPunct{\mcitedefaultmidpunct}
{\mcitedefaultendpunct}{\mcitedefaultseppunct}\relax
\EndOfBibitem
\bibitem{vanRitbergen:1998pn}
T.~van Ritbergen, A.~N. Schellekens and J.~A.~M. Vermaseren, \emph{{Group
  theory factors for Feynman diagrams}},
  \href{http://dx.doi.org/10.1142/S0217751X99000038}{\emph{Int. J. Mod. Phys.}
  {\bf A14} (1999) 41--96}, [\href{https://arxiv.org/abs/hep-ph/9802376}{{\tt
  hep-ph/9802376}}]\relax
\mciteBstWouldAddEndPuncttrue
\mciteSetBstMidEndSepPunct{\mcitedefaultmidpunct}
{\mcitedefaultendpunct}{\mcitedefaultseppunct}\relax
\EndOfBibitem
\bibitem{Lagrange:1760}
J.~L. Lagrange, \emph{{Nouvelle Recherches sur la nature et la Propagation du
  son}}, {\emph{Miscellanea Taurinensia} {\bf II} (1760-61) 11--172},
  {reprinted in Serret, J. A. (ed.), \emph{Oeuvres de Lagrange}, vol. 1, pp.
  151--316. Gauthier-Villars, Paris, 1867}\relax
\mciteBstWouldAddEndPuncttrue
\mciteSetBstMidEndSepPunct{\mcitedefaultmidpunct}
{\mcitedefaultendpunct}{\mcitedefaultseppunct}\relax
\EndOfBibitem
\bibitem{Gauss:1813}
C.~F. Gau{\ss}, \emph{{Theoria attactionis corporum sphaeroidicorum
  ellipticorum homogeneorum methodo novo tractate}}, {\emph{Commentationes
  scietas scientiarum Gottingensis recentiores} {\bf II} (1813) }, {reprinted
  in \emph{Werke}, vol. 5, pp. 3--22. G{\"o}ttingen, 1867}\relax
\mciteBstWouldAddEndPuncttrue
\mciteSetBstMidEndSepPunct{\mcitedefaultmidpunct}
{\mcitedefaultendpunct}{\mcitedefaultseppunct}\relax
\EndOfBibitem
\bibitem{Green:1828}
G.~Green, \emph{{An Essay on the Application of Mathematical Analysis to the
  Theories of Electricity and Magnetism}},  {reprinted in Ferrers, N. M. (ed.),
  \emph{Mathematical papers of the late George Green}, pp. 1--115. Macmillan,
  London, 1871}\relax
\mciteBstWouldAddEndPuncttrue
\mciteSetBstMidEndSepPunct{\mcitedefaultmidpunct}
{\mcitedefaultendpunct}{\mcitedefaultseppunct}\relax
\EndOfBibitem
\bibitem{Ostrogradski:1831}
M.~Ostrogradski, \emph{Note sur une int{\'e}grale qui se rencontre dans le
  calcul de l'attraction des sph{\'e}ro{\"i}des}, {\emph{Mem. Ac. Sci St.
  Peters.} {\bf 1} (1831) 39--53}\relax
\mciteBstWouldAddEndPuncttrue
\mciteSetBstMidEndSepPunct{\mcitedefaultmidpunct}
{\mcitedefaultendpunct}{\mcitedefaultseppunct}\relax
\EndOfBibitem
\bibitem{Chetyrkin:1980pr}
K.~G. Chetyrkin, A.~L. Kataev and F.~V. Tkachov, \emph{{New Approach to
  Evaluation of Multiloop Feynman Integrals: The Gegenbauer Polynomial x Space
  Technique}},
  \href{http://dx.doi.org/10.1016/0550-3213(80)90289-8}{\emph{Nucl. Phys.} {\bf
  B174} (1980) 345--377}\relax
\mciteBstWouldAddEndPuncttrue
\mciteSetBstMidEndSepPunct{\mcitedefaultmidpunct}
{\mcitedefaultendpunct}{\mcitedefaultseppunct}\relax
\EndOfBibitem
\bibitem{Chetyrkin:1981qh}
K.~G. Chetyrkin and F.~V. Tkachov, \emph{{Integration by Parts: The Algorithm
  to Calculate beta Functions in 4 Loops}},
  \href{http://dx.doi.org/10.1016/0550-3213(81)90199-1}{\emph{Nucl. Phys.} {\bf
  B192} (1981) 159--204}\relax
\mciteBstWouldAddEndPuncttrue
\mciteSetBstMidEndSepPunct{\mcitedefaultmidpunct}
{\mcitedefaultendpunct}{\mcitedefaultseppunct}\relax
\EndOfBibitem
\bibitem{Tkachov:1981wb}
F.~V. Tkachov, \emph{{A Theorem on Analytical Calculability of Four Loop
  Renormalization Group Functions}},
  \href{http://dx.doi.org/10.1016/0370-2693(81)90288-4}{\emph{Phys. Lett.} {\bf
  B100} (1981) 65--68}\relax
\mciteBstWouldAddEndPuncttrue
\mciteSetBstMidEndSepPunct{\mcitedefaultmidpunct}
{\mcitedefaultendpunct}{\mcitedefaultseppunct}\relax
\EndOfBibitem
\bibitem{Studerus:2009ye}
C.~Studerus, \emph{{Reduze -- Feynman Integral Reduction in C++}},
  \href{http://dx.doi.org/10.1016/j.cpc.2010.03.012}{\emph{Comput. Phys.
  Commun.} {\bf 181} (2010) 1293--1300},
  [\href{https://arxiv.org/abs/0912.2546}{{\tt 0912.2546}}]\relax
\mciteBstWouldAddEndPuncttrue
\mciteSetBstMidEndSepPunct{\mcitedefaultmidpunct}
{\mcitedefaultendpunct}{\mcitedefaultseppunct}\relax
\EndOfBibitem
\bibitem{vonManteuffel:2012np}
A.~von Manteuffel and C.~Studerus, \emph{{Reduze 2 -- Distributed Feynman
  Integral Reduction}},  \href{https://arxiv.org/abs/1201.4330}{{\tt
  1201.4330}}\relax
\mciteBstWouldAddEndPuncttrue
\mciteSetBstMidEndSepPunct{\mcitedefaultmidpunct}
{\mcitedefaultendpunct}{\mcitedefaultseppunct}\relax
\EndOfBibitem
\bibitem{Bauer:2000cp}
C.~W. Bauer, A.~Frink and R.~Kreckel, \emph{{Introduction to the GiNaC
  framework for symbolic computation within the C++ programming language}},
  \href{http://dx.doi.org/10.1006/jsco.2001.0494}{\emph{J. Symb. Comput.} {\bf
  33} (2000) 1}, [\href{https://arxiv.org/abs/cs/0004015}{{\tt
  cs/0004015}}]\relax
\mciteBstWouldAddEndPuncttrue
\mciteSetBstMidEndSepPunct{\mcitedefaultmidpunct}
{\mcitedefaultendpunct}{\mcitedefaultseppunct}\relax
\EndOfBibitem
\bibitem{Lewis:Fermat}
R.~H. Lewis, ``{Computer Algebra System Fermat}.''
\newblock \href{http://home.bway.net/lewis}{http://home.bway.net/lewis}\relax
\mciteBstWouldAddEndPuncttrue
\mciteSetBstMidEndSepPunct{\mcitedefaultmidpunct}
{\mcitedefaultendpunct}{\mcitedefaultseppunct}\relax
\EndOfBibitem
\bibitem{Gauss:1812}
C.~F. Gau{\ss}, \emph{{Disquisitiones generales circa seriam infinitam $1 +
  \tfrac{\alpha \beta}{1 \cdot \gamma} x +\text{} \tfrac{\alpha (\alpha+1)
  \beta (\beta+1)}{1 \cdot 2 \cdot \gamma (\gamma+1)} x~x +\text{etc.}$}},
  {\emph{Commentationes societatis regiae scientarum Gottingensis recentiores}
  (1812) 3--48}\relax
\mciteBstWouldAddEndPuncttrue
\mciteSetBstMidEndSepPunct{\mcitedefaultmidpunct}
{\mcitedefaultendpunct}{\mcitedefaultseppunct}\relax
\EndOfBibitem
\bibitem{Kummer:1836}
E.~E. Kummer, \emph{{{\"Uber} \hspace{0.5ex} die \hspace{0.5ex}
  hypergeometrische \hspace{0.5ex} Reihe \hspace{0.5ex} $1 + \tfrac{\alpha\cdot
  \beta}{1 \cdot \gamma} x + \tfrac{\alpha (\alpha+1) \beta (\beta+1)}{1 \cdot
  2 \cdot \gamma (\gamma+1)} x^2$ $+\text{} \tfrac{\alpha (\alpha+1)(\alpha+2)
  \beta (\beta+1)(\beta+2)}{1 \cdot 2\cdot 3 \cdot \gamma (\gamma+1)(\gamma+2)}
  x^3 +\dots$}}, {\emph{J. reine angew. Math.} {\bf 15} (1836) 39--83,
  127--172}\relax
\mciteBstWouldAddEndPuncttrue
\mciteSetBstMidEndSepPunct{\mcitedefaultmidpunct}
{\mcitedefaultendpunct}{\mcitedefaultseppunct}\relax
\EndOfBibitem
\bibitem{Bailey:1935}
W.~N. Bailey, \emph{{Generalized Hypergeometric Series}}.
\newblock No.~32 in Cambridge Tracts in Mathematics and Mathematical Physics.
  Cambridge University Press, London, 1935\relax
\mciteBstWouldAddEndPuncttrue
\mciteSetBstMidEndSepPunct{\mcitedefaultmidpunct}
{\mcitedefaultendpunct}{\mcitedefaultseppunct}\relax
\EndOfBibitem
\bibitem{Slater:1966}
L.~J. Slater, \emph{Generalized hypergeometric functions}.
\newblock Cambridge University Press, Cambridge, 1966\relax
\mciteBstWouldAddEndPuncttrue
\mciteSetBstMidEndSepPunct{\mcitedefaultmidpunct}
{\mcitedefaultendpunct}{\mcitedefaultseppunct}\relax
\EndOfBibitem
\bibitem{Gradstein:1981}
I.~S. Gradstein and I.~M. Ryshik, \emph{{Tables of series, products and
  integrals}}.
\newblock Verlag Harri Deutsch, Thun, 1981\relax
\mciteBstWouldAddEndPuncttrue
\mciteSetBstMidEndSepPunct{\mcitedefaultmidpunct}
{\mcitedefaultendpunct}{\mcitedefaultseppunct}\relax
\EndOfBibitem
\bibitem{Appell:1926}
P.~Appell and J.~Kamp{\'e}~de F{\'e}riet, \emph{{Fonctions
  hyperg{\'e}om{\'e}triques et hypersph{\'e}riques: polyn{\^o}mes d'Hermite}}.
\newblock Gauthier-Villars, Paris, 1926\relax
\mciteBstWouldAddEndPuncttrue
\mciteSetBstMidEndSepPunct{\mcitedefaultmidpunct}
{\mcitedefaultendpunct}{\mcitedefaultseppunct}\relax
\EndOfBibitem
\bibitem{Appell:1925}
P.~Appell, \emph{{Sur les fonctions hyperg{\'e}om{\'e}triques de plusieurs
  variables, les polyn{\^o}mes d'Hermite et autres fonctions sph{\'e}riques
  dans l'hyperespace}}.
\newblock Gauthier-Villars, Paris, 1925\relax
\mciteBstWouldAddEndPuncttrue
\mciteSetBstMidEndSepPunct{\mcitedefaultmidpunct}
{\mcitedefaultendpunct}{\mcitedefaultseppunct}\relax
\EndOfBibitem
\bibitem{KampeDeFeriet:1937}
J.~Kamp{\'e} De~F{\'e}riet, \emph{{La fonction hyperg{\'e}om{\'e}trique}}.
\newblock Gauthier-Villars, Paris, 1937\relax
\mciteBstWouldAddEndPuncttrue
\mciteSetBstMidEndSepPunct{\mcitedefaultmidpunct}
{\mcitedefaultendpunct}{\mcitedefaultseppunct}\relax
\EndOfBibitem
\bibitem{Exton:1976yx}
H.~Exton, \emph{{Multiple Hypergeometric Functions and Applications}}.
\newblock Ellis Horwood, Chichester, 1976\relax
\mciteBstWouldAddEndPuncttrue
\mciteSetBstMidEndSepPunct{\mcitedefaultmidpunct}
{\mcitedefaultendpunct}{\mcitedefaultseppunct}\relax
\EndOfBibitem
\bibitem{Exton:1978}
H.~Exton, \emph{{Handbook of hypergeometric integrals}}.
\newblock Ellis Horwood, Chichester, 1978\relax
\mciteBstWouldAddEndPuncttrue
\mciteSetBstMidEndSepPunct{\mcitedefaultmidpunct}
{\mcitedefaultendpunct}{\mcitedefaultseppunct}\relax
\EndOfBibitem
\bibitem{Srivastava:1985}
H.~M. Srivastava and P.~W. Karlsson, \emph{{Multiple Gaussian hypergeometric
  series}}.
\newblock Ellis Horwood, Chichester, 1985\relax
\mciteBstWouldAddEndPuncttrue
\mciteSetBstMidEndSepPunct{\mcitedefaultmidpunct}
{\mcitedefaultendpunct}{\mcitedefaultseppunct}\relax
\EndOfBibitem
\bibitem{Schneider:2001}
C.~Schneider, \emph{{Symbolic Summation in Difference Fields}}.
\newblock PhD thesis, RISC, J. Kepler University Linz, 2001\relax
\mciteBstWouldAddEndPuncttrue
\mciteSetBstMidEndSepPunct{\mcitedefaultmidpunct}
{\mcitedefaultendpunct}{\mcitedefaultseppunct}\relax
\EndOfBibitem
\bibitem{Karr:1981}
M.~Karr, \emph{{Summation in Finite Terms}},
  \href{http://dx.doi.org/10.1145/322248.322255}{\emph{J. ACM} {\bf 28} (1981)
  305--350}\relax
\mciteBstWouldAddEndPuncttrue
\mciteSetBstMidEndSepPunct{\mcitedefaultmidpunct}
{\mcitedefaultendpunct}{\mcitedefaultseppunct}\relax
\EndOfBibitem
\bibitem{Schneider:2004den}
C.~Schneider, \emph{{A Collection of Denominator Bounds to Solve Parameterized
  Linear Difference Equations in $\Pi\Sigma$-Extensions}}, {\emph{An. Univ.
  Timisoara Ser. Mat.-Inform.} {\bf 42} (2004) 163--179}\relax
\mciteBstWouldAddEndPuncttrue
\mciteSetBstMidEndSepPunct{\mcitedefaultmidpunct}
{\mcitedefaultendpunct}{\mcitedefaultseppunct}\relax
\EndOfBibitem
\bibitem{Schneider:2005rec}
C.~Schneider, \emph{{Solving Parameterized Linear Difference Equations in Terms
  of Indefinite Nested Sums and Products}},
  \href{http://dx.doi.org/10.1080/10236190500138262}{\emph{J. Differ. Equations
  Appl.} {\bf 11} (2005) 799--821}\relax
\mciteBstWouldAddEndPuncttrue
\mciteSetBstMidEndSepPunct{\mcitedefaultmidpunct}
{\mcitedefaultendpunct}{\mcitedefaultseppunct}\relax
\EndOfBibitem
\bibitem{Schneider:2005pol}
C.~Schneider, \emph{{Degree Bounds to Find Polynomial Solutions of
  Parameterized Linear Difference Equations in $\Pi\Sigma$-Fields}},
  \href{http://dx.doi.org/10.1007/s00200-004-0167-3}{\emph{Appl. Algebra Engrg.
  Comm. Comput.} {\bf 16} (2005) 1--32}\relax
\mciteBstWouldAddEndPuncttrue
\mciteSetBstMidEndSepPunct{\mcitedefaultmidpunct}
{\mcitedefaultendpunct}{\mcitedefaultseppunct}\relax
\EndOfBibitem
\bibitem{Schneider:2007sim}
C.~Schneider, \emph{{Simplifying Sums in $\Pi\Sigma^*$-Extensions}},
  \href{http://dx.doi.org/10.1142/S0219498807002302}{\emph{J. Algebra Appl.}
  {\bf 6} (2007) 415--441}\relax
\mciteBstWouldAddEndPuncttrue
\mciteSetBstMidEndSepPunct{\mcitedefaultmidpunct}
{\mcitedefaultendpunct}{\mcitedefaultseppunct}\relax
\EndOfBibitem
\bibitem{Schneider:2010cla}
C.~Schneider, \emph{{A Symbolic Summation Approach to Find Optimal Nested Sum
  Representations}},  in \emph{{Motives, Quantum Field Theory, and
  Pseudodifferential Operators}} (A.~Carey, D.~Ellwood, S.~Paycha and
  S.~Rosenberg, eds.), vol.~12 of \emph{Clay Mathematics Proceedings},
  pp.~285--308, Amer. Math. Soc, 2010.
\newblock \href{https://arxiv.org/abs/0904.2323}{{\tt 0904.2323}}\relax
\mciteBstWouldAddEndPuncttrue
\mciteSetBstMidEndSepPunct{\mcitedefaultmidpunct}
{\mcitedefaultendpunct}{\mcitedefaultseppunct}\relax
\EndOfBibitem
\bibitem{Schneider:2010alg}
C.~Schneider, \emph{{Parameterized Telescoping Proves Algebraic Independence of
  Sums}}, \href{http://dx.doi.org/10.1007/s00026-011-0076-7}{\emph{Ann. Comb.}
  {\bf 14} (2010) 533--552}, [\href{https://arxiv.org/abs/0808.2596}{{\tt
  0808.2596}}]\relax
\mciteBstWouldAddEndPuncttrue
\mciteSetBstMidEndSepPunct{\mcitedefaultmidpunct}
{\mcitedefaultendpunct}{\mcitedefaultseppunct}\relax
\EndOfBibitem
\bibitem{Schneider:2015lnc}
C.~Schneider, \emph{{Fast Algorithms for Refined Parameterized Telescoping in
  Difference Fields}},  in \emph{{Computer Algebra and Polynomials,
  Applications of Algebra and Number Theory}} (J.~Gutierrez, J.~Schicho and
  M.~Weimann, eds.), no.~8942 in Lecture Notes in Computer Science,
  pp.~157--191.
\newblock Springer, 2015.
\newblock \href{https://arxiv.org/abs/1307.7887}{{\tt 1307.7887}}\relax
\mciteBstWouldAddEndPuncttrue
\mciteSetBstMidEndSepPunct{\mcitedefaultmidpunct}
{\mcitedefaultendpunct}{\mcitedefaultseppunct}\relax
\EndOfBibitem
\bibitem{Schneider:2008jsc}
C.~Schneider, \emph{{A refined difference field theory for symbolic
  summation}}, \href{http://dx.doi.org/10.1016/j.jsc.2008.01.001}{\emph{J.
  Symbolic Comput.} {\bf 43} (2008) 611--644},
  [\href{https://arxiv.org/abs/0808.2543}{{\tt 0808.2543}}]\relax
\mciteBstWouldAddEndPuncttrue
\mciteSetBstMidEndSepPunct{\mcitedefaultmidpunct}
{\mcitedefaultendpunct}{\mcitedefaultseppunct}\relax
\EndOfBibitem
\bibitem{Schneider:2016jsc}
C.~Schneider, \emph{{A difference ring theory for symbolic summation}},
  \href{http://dx.doi.org/10.1016/j.jsc.2015.02.002}{\emph{J. Symbolic Comput.}
  {\bf 72} (2016) 82--127}, [\href{https://arxiv.org/abs/1408.2776}{{\tt
  1408.2776}}]\relax
\mciteBstWouldAddEndPuncttrue
\mciteSetBstMidEndSepPunct{\mcitedefaultmidpunct}
{\mcitedefaultendpunct}{\mcitedefaultseppunct}\relax
\EndOfBibitem
\bibitem{Schneider:2016two}
C.~Schneider, \emph{{Summation Theory II: Characterizations of
  $R\Pi\Sigma$-extensions and algorithmic aspects}},
  \href{http://dx.doi.org/10.1016/j.jsc.2016.07.028}{\emph{to appear in J.
  Symbolic Comput.} (2016) }, [\href{https://arxiv.org/abs/1603.04285}{{\tt
  1603.04285}}]\relax
\mciteBstWouldAddEndPuncttrue
\mciteSetBstMidEndSepPunct{\mcitedefaultmidpunct}
{\mcitedefaultendpunct}{\mcitedefaultseppunct}\relax
\EndOfBibitem
\bibitem{Schneider:2007}
C.~Schneider, \emph{{Symbolic Summation Assists Combinatorics}}, {\emph{Sem.
  Lothar. Combin.} {\bf 56} (2007) article 56b}\relax
\mciteBstWouldAddEndPuncttrue
\mciteSetBstMidEndSepPunct{\mcitedefaultmidpunct}
{\mcitedefaultendpunct}{\mcitedefaultseppunct}\relax
\EndOfBibitem
\bibitem{Schneider:2013book}
C.~Schneider, \emph{{Simplifying Multiple Sums in Difference Fields}},  in
  \emph{{Computer Algebra in Quantum Field Theory: Integration, Summation and
  Special Functions}} (C.~Schneider and J.~Bl{\"u}mlein, eds.), pp.~325--360.
\newblock Springer, Vienna, 2013.
\newblock \href{https://arxiv.org/abs/1304.4134}{{\tt 1304.4134}}.
\newblock \href{http://dx.doi.org/10.1007/978-3-7091-1616-6_14}{DOI}\relax
\mciteBstWouldAddEndPuncttrue
\mciteSetBstMidEndSepPunct{\mcitedefaultmidpunct}
{\mcitedefaultendpunct}{\mcitedefaultseppunct}\relax
\EndOfBibitem
\bibitem{Ablinger:2010pb}
J.~Ablinger, J.~Bl{\"u}mlein, S.~Klein and C.~Schneider, \emph{{Modern
  Summation Methods and the Computation of 2- and 3-loop Feynman Diagrams}},
  \href{http://dx.doi.org/10.1016/j.nuclphysbps.2010.08.028}{\emph{Nucl. Phys.
  Proc. Suppl.} {\bf 205-206} (2010) 110--115},
  [\href{https://arxiv.org/abs/1006.4797}{{\tt 1006.4797}}]\relax
\mciteBstWouldAddEndPuncttrue
\mciteSetBstMidEndSepPunct{\mcitedefaultmidpunct}
{\mcitedefaultendpunct}{\mcitedefaultseppunct}\relax
\EndOfBibitem
\bibitem{Blumlein:2012hg}
J.~Bl{\"u}mlein, A.~Hasselhuhn and C.~Schneider, \emph{{Evaluation of
  Multi-Sums for Large Scale Problems}}, {\emph{PoS} {\bf RADCOR2011} (2011)
  032}, [\href{https://arxiv.org/abs/1202.4303}{{\tt 1202.4303}}]\relax
\mciteBstWouldAddEndPuncttrue
\mciteSetBstMidEndSepPunct{\mcitedefaultmidpunct}
{\mcitedefaultendpunct}{\mcitedefaultseppunct}\relax
\EndOfBibitem
\bibitem{Schneider:2013}
C.~Schneider, \emph{{Symbolic Summation in Difference Fields and Its
  Application in Particle Physics}}, {\emph{Computer Algebra Rundbrief} {\bf
  53} (2013) 8--12}\relax
\mciteBstWouldAddEndPuncttrue
\mciteSetBstMidEndSepPunct{\mcitedefaultmidpunct}
{\mcitedefaultendpunct}{\mcitedefaultseppunct}\relax
\EndOfBibitem
\bibitem{Schneider:2013zna}
C.~Schneider, \emph{{Modern Summation Methods for Loop Integrals in Quantum
  Field Theory: The Packages Sigma, EvaluateMultiSums and SumProduction}},
  \href{http://dx.doi.org/10.1088/1742-6596/523/1/012037}{\emph{J. Phys. Conf.
  Ser.} {\bf 523} (2014) 012037}, [\href{https://arxiv.org/abs/1310.0160}{{\tt
  1310.0160}}]\relax
\mciteBstWouldAddEndPuncttrue
\mciteSetBstMidEndSepPunct{\mcitedefaultmidpunct}
{\mcitedefaultendpunct}{\mcitedefaultseppunct}\relax
\EndOfBibitem
\bibitem{Ablinger:2010kw}
J.~Ablinger, \emph{{A Computer Algebra Toolbox for Harmonic Sums Related to
  Particle Physics}},  {Diploma thesis}, J. Kepler University Linz, 2009\relax
\mciteBstWouldAddEndPuncttrue
\mciteSetBstMidEndSepPunct{\mcitedefaultmidpunct}
{\mcitedefaultendpunct}{\mcitedefaultseppunct}\relax
\EndOfBibitem
\bibitem{Ablinger:2013hcp}
J.~Ablinger, \emph{{Computer Algebra Algorithms for Special Functions in
  Particle Physics}}.
\newblock PhD thesis, J. Kepler University Linz, 2012.
\newblock \href{https://arxiv.org/abs/1305.0687}{{\tt 1305.0687}}\relax
\mciteBstWouldAddEndPuncttrue
\mciteSetBstMidEndSepPunct{\mcitedefaultmidpunct}
{\mcitedefaultendpunct}{\mcitedefaultseppunct}\relax
\EndOfBibitem
\bibitem{Ablinger:2011te}
J.~Ablinger, J.~Bl{\"u}mlein and C.~Schneider, \emph{{Harmonic Sums and
  Polylogarithms Generated by Cyclotomic Polynomials}},
  \href{http://dx.doi.org/10.1063/1.3629472}{\emph{J. Math. Phys.} {\bf 52}
  (2011) 102301}, [\href{https://arxiv.org/abs/1105.6063}{{\tt
  1105.6063}}]\relax
\mciteBstWouldAddEndPuncttrue
\mciteSetBstMidEndSepPunct{\mcitedefaultmidpunct}
{\mcitedefaultendpunct}{\mcitedefaultseppunct}\relax
\EndOfBibitem
\bibitem{Ablinger:2013cf}
J.~Ablinger, J.~Bl{\"u}mlein and C.~Schneider, \emph{{Analytic and Algorithmic
  Aspects of Generalized Harmonic Sums and Polylogarithms}},
  \href{http://dx.doi.org/10.1063/1.4811117}{\emph{J. Math. Phys.} {\bf 54}
  (2013) 082301}, [\href{https://arxiv.org/abs/1302.0378}{{\tt
  1302.0378}}]\relax
\mciteBstWouldAddEndPuncttrue
\mciteSetBstMidEndSepPunct{\mcitedefaultmidpunct}
{\mcitedefaultendpunct}{\mcitedefaultseppunct}\relax
\EndOfBibitem
\bibitem{Ablinger:2014bra}
J.~Ablinger, J.~Bl{\"u}mlein, C.~G. Raab and C.~Schneider, \emph{{Iterated
  Binomial Sums and their Associated Iterated Integrals}},
  \href{http://dx.doi.org/10.1063/1.4900836}{\emph{J. Math. Phys.} {\bf 55}
  (2014) 112301}, [\href{https://arxiv.org/abs/1407.1822}{{\tt
  1407.1822}}]\relax
\mciteBstWouldAddEndPuncttrue
\mciteSetBstMidEndSepPunct{\mcitedefaultmidpunct}
{\mcitedefaultendpunct}{\mcitedefaultseppunct}\relax
\EndOfBibitem
\bibitem{Ablinger:2014rba}
J.~Ablinger, \emph{{The package HarmonicSums: Computer Algebra and Analytic
  aspects of Nested Sums}}, {\emph{PoS} {\bf LL2014} (2014) 019},
  [\href{https://arxiv.org/abs/1407.6180}{{\tt 1407.6180}}]\relax
\mciteBstWouldAddEndPuncttrue
\mciteSetBstMidEndSepPunct{\mcitedefaultmidpunct}
{\mcitedefaultendpunct}{\mcitedefaultseppunct}\relax
\EndOfBibitem
\bibitem{Mellin:1895}
H.~Mellin, \emph{{Om definita integraler, hvilka f{\"o}r obegr{\"a}nsadt
  v{\"a}xande v{\"a}rden af vissa heltaliga parametrar hafva till gr{\"a}nser
  hypergeometriska funktioner af s{\"a}rskilda ordningar}}, {\emph{Acta
  Societatis Scientiarum Fennicae} {\bf XX} (1885) 1--39}\relax
\mciteBstWouldAddEndPuncttrue
\mciteSetBstMidEndSepPunct{\mcitedefaultmidpunct}
{\mcitedefaultendpunct}{\mcitedefaultseppunct}\relax
\EndOfBibitem
\bibitem{Barnes:1908}
E.~W. Barnes, \emph{{A New Development of the Theory of the Hypergeometric
  Functions}}, \href{http://dx.doi.org/10.1112/plms/s2-6.1.141}{\emph{Proc.
  London Math. Soc.} {\bf s2-6} (1908) 141--177}\relax
\mciteBstWouldAddEndPuncttrue
\mciteSetBstMidEndSepPunct{\mcitedefaultmidpunct}
{\mcitedefaultendpunct}{\mcitedefaultseppunct}\relax
\EndOfBibitem
\bibitem{Barnes:1910}
E.~W. Barnes, \emph{{A Transformation of Generalised Hypergeometric Series}},
  {\emph{Quart. J. Math.} {\bf 41} (1910) 136--140}\relax
\mciteBstWouldAddEndPuncttrue
\mciteSetBstMidEndSepPunct{\mcitedefaultmidpunct}
{\mcitedefaultendpunct}{\mcitedefaultseppunct}\relax
\EndOfBibitem
\bibitem{Smirnov:1999gc}
V.~A. Smirnov, \emph{{Analytical result for dimensionally regularized massless
  on shell double box}},
  \href{http://dx.doi.org/10.1016/S0370-2693(99)00777-7}{\emph{Phys. Lett.}
  {\bf B460} (1999) 397--404},
  [\href{https://arxiv.org/abs/hep-ph/9905323}{{\tt hep-ph/9905323}}]\relax
\mciteBstWouldAddEndPuncttrue
\mciteSetBstMidEndSepPunct{\mcitedefaultmidpunct}
{\mcitedefaultendpunct}{\mcitedefaultseppunct}\relax
\EndOfBibitem
\bibitem{Tausk:1999vh}
J.~B. Tausk, \emph{{Nonplanar massless two loop Feynman diagrams with four
  on-shell legs}},
  \href{http://dx.doi.org/10.1016/S0370-2693(99)01277-0}{\emph{Phys. Lett.}
  {\bf B469} (1999) 225--234},
  [\href{https://arxiv.org/abs/hep-ph/9909506}{{\tt hep-ph/9909506}}]\relax
\mciteBstWouldAddEndPuncttrue
\mciteSetBstMidEndSepPunct{\mcitedefaultmidpunct}
{\mcitedefaultendpunct}{\mcitedefaultseppunct}\relax
\EndOfBibitem
\bibitem{Smirnov:2006ry}
V.~A. Smirnov, \emph{{Feynman integral calculus}}.
\newblock Springer, Berlin, 2006\relax
\mciteBstWouldAddEndPuncttrue
\mciteSetBstMidEndSepPunct{\mcitedefaultmidpunct}
{\mcitedefaultendpunct}{\mcitedefaultseppunct}\relax
\EndOfBibitem
\bibitem{Czakon:2005rk}
M.~Czakon, \emph{{Automatized analytic continuation of Mellin-Barnes
  integrals}}, \href{http://dx.doi.org/10.1016/j.cpc.2006.07.002}{\emph{Comput.
  Phys. Commun.} {\bf 175} (2006) 559--571},
  [\href{https://arxiv.org/abs/hep-ph/0511200}{{\tt hep-ph/0511200}}]\relax
\mciteBstWouldAddEndPuncttrue
\mciteSetBstMidEndSepPunct{\mcitedefaultmidpunct}
{\mcitedefaultendpunct}{\mcitedefaultseppunct}\relax
\EndOfBibitem
\bibitem{Smirnov:2009up}
A.~V. Smirnov and V.~A. Smirnov, \emph{{On the Resolution of Singularities of
  Multiple Mellin-Barnes Integrals}},
  \href{http://dx.doi.org/10.1140/epjc/s10052-009-1039-6}{\emph{Eur. Phys. J.}
  {\bf C62} (2009) 445--449}, [\href{https://arxiv.org/abs/0901.0386}{{\tt
  0901.0386}}]\relax
\mciteBstWouldAddEndPuncttrue
\mciteSetBstMidEndSepPunct{\mcitedefaultmidpunct}
{\mcitedefaultendpunct}{\mcitedefaultseppunct}\relax
\EndOfBibitem
\bibitem{Almkvist:1990}
G.~Almkvist and D.~Zeilberger, \emph{{The method of differentiating under the
  integral sign}},
  \href{http://dx.doi.org/10.1016/S0747-7171(08)80159-9}{\emph{J. Symb. Comp.}
  {\bf 10} (1990) 571 -- 591}\relax
\mciteBstWouldAddEndPuncttrue
\mciteSetBstMidEndSepPunct{\mcitedefaultmidpunct}
{\mcitedefaultendpunct}{\mcitedefaultseppunct}\relax
\EndOfBibitem
\bibitem{Apagodu:2006}
M.~Apagodu and D.~Zeilberger, \emph{{Multi-variable Zeilberger and
  Almkvist--Zeilberger algorithms and the sharpening of Wilf--Zeilberger
  theory}}, \href{http://dx.doi.org/10.1016/j.aam.2005.09.003}{\emph{Adv. Appl.
  Math.} {\bf 37} (2006) 139 -- 152}\relax
\mciteBstWouldAddEndPuncttrue
\mciteSetBstMidEndSepPunct{\mcitedefaultmidpunct}
{\mcitedefaultendpunct}{\mcitedefaultseppunct}\relax
\EndOfBibitem
\bibitem{Kotikov:1990kg}
A.~V. Kotikov, \emph{{Differential equations method: New technique for massive
  Feynman diagrams calculation}},
  \href{http://dx.doi.org/10.1016/0370-2693(91)90413-K}{\emph{Phys. Lett.} {\bf
  B254} (1991) 158--164}\relax
\mciteBstWouldAddEndPuncttrue
\mciteSetBstMidEndSepPunct{\mcitedefaultmidpunct}
{\mcitedefaultendpunct}{\mcitedefaultseppunct}\relax
\EndOfBibitem
\bibitem{Caffo:1998yd}
M.~Caffo, H.~Czy{\.z}, S.~Laporta and E.~Remiddi, \emph{{Master equations for
  master amplitudes}}, {\emph{Acta Phys. Polon.} {\bf B29} (1998) 2627--2635},
  [\href{https://arxiv.org/abs/hep-th/9807119}{{\tt hep-th/9807119}}]\relax
\mciteBstWouldAddEndPuncttrue
\mciteSetBstMidEndSepPunct{\mcitedefaultmidpunct}
{\mcitedefaultendpunct}{\mcitedefaultseppunct}\relax
\EndOfBibitem
\bibitem{Caffo:1998du}
M.~Caffo, H.~Czy{\.z}, S.~Laporta and E.~Remiddi, \emph{{The Master
  differential equations for the two loop sunrise selfmass amplitudes}},
  {\emph{Nuovo Cim.} {\bf A111} (1998) 365--389},
  [\href{https://arxiv.org/abs/hep-th/9805118}{{\tt hep-th/9805118}}]\relax
\mciteBstWouldAddEndPuncttrue
\mciteSetBstMidEndSepPunct{\mcitedefaultmidpunct}
{\mcitedefaultendpunct}{\mcitedefaultseppunct}\relax
\EndOfBibitem
\bibitem{Gehrmann:1999as}
T.~Gehrmann and E.~Remiddi, \emph{{Differential equations for two loop four
  point functions}},
  \href{http://dx.doi.org/10.1016/S0550-3213(00)00223-6}{\emph{Nucl. Phys.}
  {\bf B580} (2000) 485--518},
  [\href{https://arxiv.org/abs/hep-ph/9912329}{{\tt hep-ph/9912329}}]\relax
\mciteBstWouldAddEndPuncttrue
\mciteSetBstMidEndSepPunct{\mcitedefaultmidpunct}
{\mcitedefaultendpunct}{\mcitedefaultseppunct}\relax
\EndOfBibitem
\bibitem{Ablinger:2015tua}
J.~Ablinger, A.~Behring, J.~Bl{\"u}mlein, A.~De~Freitas, A.~von Manteuffel and
  C.~Schneider, \emph{{Calculating Three Loop Ladder and V-Topologies for
  Massive Operator Matrix Elements by Computer Algebra}},
  \href{http://dx.doi.org/10.1016/j.cpc.2016.01.002}{\emph{Comput. Phys.
  Commun.} {\bf 202} (2016) 33--112},
  [\href{https://arxiv.org/abs/1509.08324}{{\tt 1509.08324}}]\relax
\mciteBstWouldAddEndPuncttrue
\mciteSetBstMidEndSepPunct{\mcitedefaultmidpunct}
{\mcitedefaultendpunct}{\mcitedefaultseppunct}\relax
\EndOfBibitem
\bibitem{Zuercher:1994}
B.~Z{\"u}rcher, \emph{{Rationale Normalformen von pseudo-linearen
  Abbildungen}},  {Diploma thesis}, ETH Z{\"u}rich, 1994\relax
\mciteBstWouldAddEndPuncttrue
\mciteSetBstMidEndSepPunct{\mcitedefaultmidpunct}
{\mcitedefaultendpunct}{\mcitedefaultseppunct}\relax
\EndOfBibitem
\bibitem{Gerhold:2002}
S.~Gerhold, \emph{{Uncoupling Systems of Linear Ore Operator Equations}},
  {Diploma thesis}, RISC, J. Kepler University Linz, 2002\relax
\mciteBstWouldAddEndPuncttrue
\mciteSetBstMidEndSepPunct{\mcitedefaultmidpunct}
{\mcitedefaultendpunct}{\mcitedefaultseppunct}\relax
\EndOfBibitem
\bibitem{Ablinger:2016pbw}
J.~Ablinger, J.~Bluemlein, A.~de~Freitas and C.~Schneider, \emph{{A toolbox to
  solve coupled systems of differential and difference equations}}, {\emph{PoS}
  {\bf RADCOR2015} (2015) 060}, [\href{https://arxiv.org/abs/1601.01856}{{\tt
  1601.01856}}]\relax
\mciteBstWouldAddEndPuncttrue
\mciteSetBstMidEndSepPunct{\mcitedefaultmidpunct}
{\mcitedefaultendpunct}{\mcitedefaultseppunct}\relax
\EndOfBibitem
\bibitem{Ablinger:2016yjz}
J.~Ablinger, A.~Behring, J.~Bluemlein, A.~de~Freitas and C.~Schneider,
  \emph{{Algorithms to solve coupled systems of differential equations in terms
  of power series}}, {\emph{PoS} {\bf LL2016} (2016) 005},
  [\href{https://arxiv.org/abs/1608.05376}{{\tt 1608.05376}}]\relax
\mciteBstWouldAddEndPuncttrue
\mciteSetBstMidEndSepPunct{\mcitedefaultmidpunct}
{\mcitedefaultendpunct}{\mcitedefaultseppunct}\relax
\EndOfBibitem
\bibitem{Ablinger:2010ty}
J.~Ablinger, J.~Bl{\"u}mlein, S.~Klein, C.~Schneider and F.~Wi\ss{}brock,
  \emph{{The $O(\alpha_s^3)$ Massive Operator Matrix Elements of $O(n_f)$ for
  the Structure Function $F_2(x,Q^2)$ and Transversity}},
  \href{http://dx.doi.org/10.1016/j.nuclphysb.2010.10.021}{\emph{Nucl. Phys.}
  {\bf B844} (2011) 26--54}, [\href{https://arxiv.org/abs/1008.3347}{{\tt
  1008.3347}}]\relax
\mciteBstWouldAddEndPuncttrue
\mciteSetBstMidEndSepPunct{\mcitedefaultmidpunct}
{\mcitedefaultendpunct}{\mcitedefaultseppunct}\relax
\EndOfBibitem
\bibitem{Blumlein:2012vq}
J.~Bl{\"u}mlein, A.~Hasselhuhn, S.~Klein and C.~Schneider, \emph{{The
  $O(\alpha_s^3 n_f T_F^2 C_{A,F})$ Contributions to the Gluonic Massive
  Operator Matrix Elements}},
  \href{http://dx.doi.org/10.1016/j.nuclphysb.2012.09.001}{\emph{Nucl. Phys.}
  {\bf B866} (2013) 196--211}, [\href{https://arxiv.org/abs/1205.4184}{{\tt
  1205.4184}}]\relax
\mciteBstWouldAddEndPuncttrue
\mciteSetBstMidEndSepPunct{\mcitedefaultmidpunct}
{\mcitedefaultendpunct}{\mcitedefaultseppunct}\relax
\EndOfBibitem
\bibitem{Ablinger:2012qm}
J.~Ablinger, J.~Bl{\"u}mlein, A.~Hasselhuhn, S.~Klein, C.~Schneider and
  F.~Wi\ss{}brock, \emph{{Massive 3-loop Ladder Diagrams for Quarkonic Local
  Operator Matrix Elements}},
  \href{http://dx.doi.org/10.1016/j.nuclphysb.2012.06.007}{\emph{Nucl. Phys.}
  {\bf B864} (2012) 52--84}, [\href{https://arxiv.org/abs/1206.2252}{{\tt
  1206.2252}}]\relax
\mciteBstWouldAddEndPuncttrue
\mciteSetBstMidEndSepPunct{\mcitedefaultmidpunct}
{\mcitedefaultendpunct}{\mcitedefaultseppunct}\relax
\EndOfBibitem
\bibitem{Ablinger:2014lka}
J.~Ablinger, J.~Bl{\"u}mlein, A.~De~Freitas, A.~Hasselhuhn, A.~von Manteuffel,
  M.~Round et~al., \emph{{The Transition Matrix Element $A_{gq}(N)$ of the
  Variable Flavor Number Scheme at $O(\alpha_s^3)$}},
  \href{http://dx.doi.org/10.1016/j.nuclphysb.2014.02.007}{\emph{Nucl. Phys.}
  {\bf B882} (2014) 263--288}, [\href{https://arxiv.org/abs/1402.0359}{{\tt
  1402.0359}}]\relax
\mciteBstWouldAddEndPuncttrue
\mciteSetBstMidEndSepPunct{\mcitedefaultmidpunct}
{\mcitedefaultendpunct}{\mcitedefaultseppunct}\relax
\EndOfBibitem
\bibitem{Ablinger:2014uka}
J.~Ablinger, J.~Bl{\"u}mlein, A.~De~Freitas, A.~Hasselhuhn, A.~von Manteuffel,
  M.~Round et~al., \emph{{The $O(\alpha_s^3 T_F^2)$ Contributions to the
  Gluonic Operator Matrix Element}},
  \href{http://dx.doi.org/10.1016/j.nuclphysb.2014.05.028}{\emph{Nucl. Phys.}
  {\bf B885} (2014) 280--317}, [\href{https://arxiv.org/abs/1405.4259}{{\tt
  1405.4259}}]\relax
\mciteBstWouldAddEndPuncttrue
\mciteSetBstMidEndSepPunct{\mcitedefaultmidpunct}
{\mcitedefaultendpunct}{\mcitedefaultseppunct}\relax
\EndOfBibitem
\bibitem{Ablinger:2014yaa}
J.~Ablinger, J.~Bl{\"u}mlein, C.~Raab, C.~Schneider and F.~Wi{\ss}brock,
  \emph{{Calculating Massive 3-loop Graphs for Operator Matrix Elements by the
  Method of Hyperlogarithms}},
  \href{http://dx.doi.org/10.1016/j.nuclphysb.2014.04.007}{\emph{Nucl. Phys.}
  {\bf B885} (2014) 409--447}, [\href{https://arxiv.org/abs/1403.1137}{{\tt
  1403.1137}}]\relax
\mciteBstWouldAddEndPuncttrue
\mciteSetBstMidEndSepPunct{\mcitedefaultmidpunct}
{\mcitedefaultendpunct}{\mcitedefaultseppunct}\relax
\EndOfBibitem
\bibitem{Vermaseren:1998uu}
J.~A.~M. Vermaseren, \emph{{Harmonic sums, Mellin transforms and integrals}},
  \href{http://dx.doi.org/10.1142/S0217751X99001032}{\emph{Int. J. Mod. Phys.}
  {\bf A14} (1999) 2037--2076},
  [\href{https://arxiv.org/abs/hep-ph/9806280}{{\tt hep-ph/9806280}}]\relax
\mciteBstWouldAddEndPuncttrue
\mciteSetBstMidEndSepPunct{\mcitedefaultmidpunct}
{\mcitedefaultendpunct}{\mcitedefaultseppunct}\relax
\EndOfBibitem
\bibitem{Blumlein:1998if}
J.~Bl{\"u}mlein and S.~Kurth, \emph{{Harmonic sums and Mellin transforms up to
  two loop order}},
  \href{http://dx.doi.org/10.1103/PhysRevD.60.014018}{\emph{Phys. Rev.} {\bf
  D60} (1999) 014018}, [\href{https://arxiv.org/abs/hep-ph/9810241}{{\tt
  hep-ph/9810241}}]\relax
\mciteBstWouldAddEndPuncttrue
\mciteSetBstMidEndSepPunct{\mcitedefaultmidpunct}
{\mcitedefaultendpunct}{\mcitedefaultseppunct}\relax
\EndOfBibitem
\bibitem{Blumlein:2000hw}
J.~Bl{\"u}mlein, \emph{{Analytic continuation of Mellin transforms up to two
  loop order}},
  \href{http://dx.doi.org/10.1016/S0010-4655(00)00156-9}{\emph{Comput. Phys.
  Commun.} {\bf 133} (2000) 76--104},
  [\href{https://arxiv.org/abs/hep-ph/0003100}{{\tt hep-ph/0003100}}]\relax
\mciteBstWouldAddEndPuncttrue
\mciteSetBstMidEndSepPunct{\mcitedefaultmidpunct}
{\mcitedefaultendpunct}{\mcitedefaultseppunct}\relax
\EndOfBibitem
\bibitem{Blumlein:2003gb}
J.~Bl{\"u}mlein, \emph{{Algebraic relations between harmonic sums and
  associated quantities}},
  \href{http://dx.doi.org/10.1016/j.cpc.2003.12.004}{\emph{Comput. Phys.
  Commun.} {\bf 159} (2004) 19--54},
  [\href{https://arxiv.org/abs/hep-ph/0311046}{{\tt hep-ph/0311046}}]\relax
\mciteBstWouldAddEndPuncttrue
\mciteSetBstMidEndSepPunct{\mcitedefaultmidpunct}
{\mcitedefaultendpunct}{\mcitedefaultseppunct}\relax
\EndOfBibitem
\bibitem{Blumlein:2009ta}
J.~Bl{\"u}mlein, \emph{{Structural Relations of Harmonic Sums and Mellin
  Transforms up to Weight $w = 5$}},
  \href{http://dx.doi.org/10.1016/j.cpc.2009.07.004}{\emph{Comput. Phys.
  Commun.} {\bf 180} (2009) 2218--2249},
  [\href{https://arxiv.org/abs/0901.3106}{{\tt 0901.3106}}]\relax
\mciteBstWouldAddEndPuncttrue
\mciteSetBstMidEndSepPunct{\mcitedefaultmidpunct}
{\mcitedefaultendpunct}{\mcitedefaultseppunct}\relax
\EndOfBibitem
\bibitem{Blumlein:2009fz}
J.~Bl{\"u}mlein, \emph{{Structural Relations of Harmonic Sums and Mellin
  Transforms at Weight $w = 6$}},  in \emph{{Motives, Quantum Field Theory, and
  Pseudodifferential Operators}} (A.~Carey, D.~Ellwood, S.~Paycha and
  S.~Rosenberg, eds.), vol.~12 of \emph{Clay Mathematics Proceedings}, p.~167,
  Amer. Math. Soc, 2010.
\newblock \href{https://arxiv.org/abs/0901.0837}{{\tt 0901.0837}}\relax
\mciteBstWouldAddEndPuncttrue
\mciteSetBstMidEndSepPunct{\mcitedefaultmidpunct}
{\mcitedefaultendpunct}{\mcitedefaultseppunct}\relax
\EndOfBibitem
\bibitem{Moch:2001zr}
S.~Moch, P.~Uwer and S.~Weinzierl, \emph{{Nested sums, expansion of
  transcendental functions and multiscale multiloop integrals}},
  \href{http://dx.doi.org/10.1063/1.1471366}{\emph{J. Math. Phys.} {\bf 43}
  (2002) 3363--3386}, [\href{https://arxiv.org/abs/hep-ph/0110083}{{\tt
  hep-ph/0110083}}]\relax
\mciteBstWouldAddEndPuncttrue
\mciteSetBstMidEndSepPunct{\mcitedefaultmidpunct}
{\mcitedefaultendpunct}{\mcitedefaultseppunct}\relax
\EndOfBibitem
\bibitem{Fleischer:1998nb}
J.~Fleischer, A.~V. Kotikov and O.~L. Veretin, \emph{{Analytic two loop results
  for selfenergy type and vertex type diagrams with one nonzero mass}},
  \href{http://dx.doi.org/10.1016/S0550-3213(99)00078-4}{\emph{Nucl. Phys.}
  {\bf B547} (1999) 343--374},
  [\href{https://arxiv.org/abs/hep-ph/9808242}{{\tt hep-ph/9808242}}]\relax
\mciteBstWouldAddEndPuncttrue
\mciteSetBstMidEndSepPunct{\mcitedefaultmidpunct}
{\mcitedefaultendpunct}{\mcitedefaultseppunct}\relax
\EndOfBibitem
\bibitem{Davydychev:2003mv}
A.~I. Davydychev and M.~Y. Kalmykov, \emph{{Massive Feynman diagrams and
  inverse binomial sums}},
  \href{http://dx.doi.org/10.1016/j.nuclphysb.2004.08.020}{\emph{Nucl. Phys.}
  {\bf B699} (2004) 3--64}, [\href{https://arxiv.org/abs/hep-th/0303162}{{\tt
  hep-th/0303162}}]\relax
\mciteBstWouldAddEndPuncttrue
\mciteSetBstMidEndSepPunct{\mcitedefaultmidpunct}
{\mcitedefaultendpunct}{\mcitedefaultseppunct}\relax
\EndOfBibitem
\bibitem{Weinzierl:2004bn}
S.~Weinzierl, \emph{{Expansion around half integer values, binomial sums and
  inverse binomial sums}}, \href{http://dx.doi.org/10.1063/1.1758319}{\emph{J.
  Math. Phys.} {\bf 45} (2004) 2656--2673},
  [\href{https://arxiv.org/abs/hep-ph/0402131}{{\tt hep-ph/0402131}}]\relax
\mciteBstWouldAddEndPuncttrue
\mciteSetBstMidEndSepPunct{\mcitedefaultmidpunct}
{\mcitedefaultendpunct}{\mcitedefaultseppunct}\relax
\EndOfBibitem
\bibitem{Remiddi:1999ew}
E.~Remiddi and J.~A.~M. Vermaseren, \emph{{Harmonic polylogarithms}},
  \href{http://dx.doi.org/10.1142/S0217751X00000367}{\emph{Int. J. Mod. Phys.}
  {\bf A15} (2000) 725--754}, [\href{https://arxiv.org/abs/hep-ph/9905237}{{\tt
  hep-ph/9905237}}]\relax
\mciteBstWouldAddEndPuncttrue
\mciteSetBstMidEndSepPunct{\mcitedefaultmidpunct}
{\mcitedefaultendpunct}{\mcitedefaultseppunct}\relax
\EndOfBibitem
\bibitem{Ablinger:2014nga}
J.~Ablinger, A.~Behring, J.~Bl{\"u}mlein, A.~De~Freitas, A.~von Manteuffel and
  C.~Schneider, \emph{{The 3-loop pure singlet heavy flavor contributions to
  the structure function $F_2(x,Q^2)$ and the anomalous dimension}},
  \href{http://dx.doi.org/10.1016/j.nuclphysb.2014.10.008}{\emph{Nucl. Phys.}
  {\bf B890} (2014) 48--151}, [\href{https://arxiv.org/abs/1409.1135}{{\tt
  1409.1135}}]\relax
\mciteBstWouldAddEndPuncttrue
\mciteSetBstMidEndSepPunct{\mcitedefaultmidpunct}
{\mcitedefaultendpunct}{\mcitedefaultseppunct}\relax
\EndOfBibitem
\bibitem{Ablinger:2014vwa}
J.~Ablinger, A.~Behring, J.~Bl{\"u}mlein, A.~De~Freitas, A.~Hasselhuhn, A.~von
  Manteuffel et~al., \emph{{The 3-Loop Non-Singlet Heavy Flavor Contributions
  and Anomalous Dimensions for the Structure Function $F_2(x,Q^2)$ and
  Transversity}},
  \href{http://dx.doi.org/10.1016/j.nuclphysb.2014.07.010}{\emph{Nucl. Phys.}
  {\bf B886} (2014) 733--823}, [\href{https://arxiv.org/abs/1406.4654}{{\tt
  1406.4654}}]\relax
\mciteBstWouldAddEndPuncttrue
\mciteSetBstMidEndSepPunct{\mcitedefaultmidpunct}
{\mcitedefaultendpunct}{\mcitedefaultseppunct}\relax
\EndOfBibitem
\bibitem{DESY-15–112}
J.~Ablinger et~al. DESY-15-112\relax
\mciteBstWouldAddEndPuncttrue
\mciteSetBstMidEndSepPunct{\mcitedefaultmidpunct}
{\mcitedefaultendpunct}{\mcitedefaultseppunct}\relax
\EndOfBibitem
\bibitem{Blumlein:2010rn}
J.~Bl{\"u}mlein and H.~B{\"o}ttcher, \emph{{QCD Analysis of Polarized Deep
  Inelastic Scattering Data}},
  \href{http://dx.doi.org/10.1016/j.nuclphysb.2010.08.005}{\emph{Nucl. Phys.}
  {\bf B841} (2010) 205--230}, [\href{https://arxiv.org/abs/1005.3113}{{\tt
  1005.3113}}]\relax
\mciteBstWouldAddEndPuncttrue
\mciteSetBstMidEndSepPunct{\mcitedefaultmidpunct}
{\mcitedefaultendpunct}{\mcitedefaultseppunct}\relax
\EndOfBibitem
\bibitem{Behring:2015zaa}
A.~Behring, J.~Bl{\"u}mlein, A.~De~Freitas, A.~von Manteuffel and C.~Schneider,
  \emph{{The 3-Loop Non-Singlet Heavy Flavor Contributions to the Structure
  Function $g_1(x,Q^2)$ at Large Momentum Transfer}},
  \href{http://dx.doi.org/10.1016/j.nuclphysb.2015.06.007}{\emph{Nucl. Phys.}
  {\bf B897} (2015) 612--644}, [\href{https://arxiv.org/abs/1504.08217}{{\tt
  1504.08217}}]\relax
\mciteBstWouldAddEndPuncttrue
\mciteSetBstMidEndSepPunct{\mcitedefaultmidpunct}
{\mcitedefaultendpunct}{\mcitedefaultseppunct}\relax
\EndOfBibitem
\bibitem{Boer:2011fh}
D.~Boer et~al., \emph{{Gluons and the quark sea at high energies:
  Distributions, polarization, tomography}},
  \href{https://arxiv.org/abs/1108.1713}{{\tt 1108.1713}}\relax
\mciteBstWouldAddEndPuncttrue
\mciteSetBstMidEndSepPunct{\mcitedefaultmidpunct}
{\mcitedefaultendpunct}{\mcitedefaultseppunct}\relax
\EndOfBibitem
\bibitem{Accardi:2012qut}
A.~Accardi et~al., \emph{{Electron Ion Collider: The Next QCD Frontier -
  Understanding the glue that binds us all}},
  \href{https://arxiv.org/abs/1212.1701}{{\tt 1212.1701}}\relax
\mciteBstWouldAddEndPuncttrue
\mciteSetBstMidEndSepPunct{\mcitedefaultmidpunct}
{\mcitedefaultendpunct}{\mcitedefaultseppunct}\relax
\EndOfBibitem
\bibitem{AbelleiraFernandez:2012cc}
{\scshape LHeC Study Group} collaboration, J.~L. Abelleira~Fernandez et~al.,
  \emph{{A Large Hadron Electron Collider at CERN: Report on the Physics and
  Design Concepts for Machine and Detector}},
  \href{http://dx.doi.org/10.1088/0954-3899/39/7/075001}{\emph{J. Phys.} {\bf
  G39} (2012) 075001}, [\href{https://arxiv.org/abs/1206.2913}{{\tt
  1206.2913}}]\relax
\mciteBstWouldAddEndPuncttrue
\mciteSetBstMidEndSepPunct{\mcitedefaultmidpunct}
{\mcitedefaultendpunct}{\mcitedefaultseppunct}\relax
\EndOfBibitem
\bibitem{Wandzura:1977qf}
S.~Wandzura and F.~Wilczek, \emph{{Sum Rules for Spin Dependent
  Electroproduction: Test of Relativistic Constituent Quarks}},
  \href{http://dx.doi.org/10.1016/0370-2693(77)90700-6}{\emph{Phys. Lett.} {\bf
  B72} (1977) 195}\relax
\mciteBstWouldAddEndPuncttrue
\mciteSetBstMidEndSepPunct{\mcitedefaultmidpunct}
{\mcitedefaultendpunct}{\mcitedefaultseppunct}\relax
\EndOfBibitem
\bibitem{Bjorken:1969mm}
J.~D. Bjorken, \emph{{Inelastic Scattering of Polarized Leptons from Polarized
  Nucleons}}, \href{http://dx.doi.org/10.1103/PhysRevD.1.1376}{\emph{Phys.
  Rev.} {\bf D1} (1970) 1376--1379}\relax
\mciteBstWouldAddEndPuncttrue
\mciteSetBstMidEndSepPunct{\mcitedefaultmidpunct}
{\mcitedefaultendpunct}{\mcitedefaultseppunct}\relax
\EndOfBibitem
\bibitem{Baikov:2010je}
P.~A. Baikov, K.~G. Chetyrkin and J.~H. K{\"u}hn, \emph{{Adler Function,
  Bjorken Sum Rule, and the Crewther Relation to Order $\alpha_s^4$ in a
  General Gauge Theory}},
  \href{http://dx.doi.org/10.1103/PhysRevLett.104.132004}{\emph{Phys. Rev.
  Lett.} {\bf 104} (2010) 132004}, [\href{https://arxiv.org/abs/1001.3606}{{\tt
  1001.3606}}]\relax
\mciteBstWouldAddEndPuncttrue
\mciteSetBstMidEndSepPunct{\mcitedefaultmidpunct}
{\mcitedefaultendpunct}{\mcitedefaultseppunct}\relax
\EndOfBibitem
\bibitem{Larin:2013yba}
S.~A. Larin, \emph{{The singlet contribution to the Bjorken sum rule for
  polarized deep inelastic scattering}},
  \href{http://dx.doi.org/10.1016/j.physletb.2013.05.026}{\emph{Phys. Lett.}
  {\bf B723} (2013) 348--350}, [\href{https://arxiv.org/abs/1303.4021}{{\tt
  1303.4021}}]\relax
\mciteBstWouldAddEndPuncttrue
\mciteSetBstMidEndSepPunct{\mcitedefaultmidpunct}
{\mcitedefaultendpunct}{\mcitedefaultseppunct}\relax
\EndOfBibitem
\bibitem{Baikov:2015tea}
P.~A. Baikov, K.~G. Chetyrkin and J.~H. K{\"u}hn, \emph{{Massless Propagators,
  $R(s)$ and Multiloop QCD}},
  \href{http://dx.doi.org/10.1016/j.nuclphysbps.2015.03.002}{\emph{Nucl. Part.
  Phys. Proc.} {\bf 261-262} (2015) 3--18},
  [\href{https://arxiv.org/abs/1501.06739}{{\tt 1501.06739}}]\relax
\mciteBstWouldAddEndPuncttrue
\mciteSetBstMidEndSepPunct{\mcitedefaultmidpunct}
{\mcitedefaultendpunct}{\mcitedefaultseppunct}\relax
\EndOfBibitem
\bibitem{Blumlein:2016xcy}
J.~Bl{\"u}mlein, G.~Falcioni and A.~De~Freitas, \emph{{The Complete
  $O(\alpha_s^2)$ Non-Singlet Heavy Flavor Corrections to the Structure
  Functions $g_{1,2}^{ep}(x,Q^2)$, $F_{1,2,L}^{ep}(x,Q^2)$,
  $F_{1,2,3}^{\nu(\bar{\nu})}(x,Q^2)$ and the Associated Sum Rules}},
  \href{http://dx.doi.org/10.1016/j.nuclphysb.2016.06.018}{\emph{Nucl. Phys.}
  {\bf B910} (2016) 568--617}, [\href{https://arxiv.org/abs/1605.05541}{{\tt
  1605.05541}}]\relax
\mciteBstWouldAddEndPuncttrue
\mciteSetBstMidEndSepPunct{\mcitedefaultmidpunct}
{\mcitedefaultendpunct}{\mcitedefaultseppunct}\relax
\EndOfBibitem
\bibitem{Behring:2015roa}
A.~Behring, J.~Bl{\"u}mlein, A.~De~Freitas, A.~Hasselhuhn, A.~von Manteuffel
  and C.~Schneider, \emph{{$O(\alpha_s^3)$ heavy flavor contributions to the
  charged current structure function $x F_3(x,Q^2)$ at large momentum
  transfer}}, \href{http://dx.doi.org/10.1103/PhysRevD.92.114005}{\emph{Phys.
  Rev.} {\bf D92} (2015) 114005}, [\href{https://arxiv.org/abs/1508.01449}{{\tt
  1508.01449}}]\relax
\mciteBstWouldAddEndPuncttrue
\mciteSetBstMidEndSepPunct{\mcitedefaultmidpunct}
{\mcitedefaultendpunct}{\mcitedefaultseppunct}\relax
\EndOfBibitem
\bibitem{Blumlein:2011zu}
J.~Bl{\"u}mlein, A.~Hasselhuhn, P.~Kova\v{c}\'{i}kov\'{a} and S.~Moch,
  \emph{{$O(\alpha_s)$ Heavy Flavor Corrections to Charged Current
  Deep-Inelastic Scattering in Mellin Space}},
  \href{http://dx.doi.org/10.1016/j.physletb.2011.05.007}{\emph{Phys. Lett.}
  {\bf B700} (2011) 294--304}, [\href{https://arxiv.org/abs/1104.3449}{{\tt
  1104.3449}}]\relax
\mciteBstWouldAddEndPuncttrue
\mciteSetBstMidEndSepPunct{\mcitedefaultmidpunct}
{\mcitedefaultendpunct}{\mcitedefaultseppunct}\relax
\EndOfBibitem
\bibitem{Blumlein:2014fqa}
J.~Bl{\"u}mlein, A.~Hasselhuhn and T.~Pfoh, \emph{{The $O(\alpha_s^2)$ heavy
  quark corrections to charged current {} deep-inelastic {} scattering {} at {}
  large {} virtualities}},
  \href{http://dx.doi.org/10.1016/j.nuclphysb.2014.01.023}{\emph{Nucl. Phys.}
  {\bf B881} (2014) 1--41}, [\href{https://arxiv.org/abs/1401.4352}{{\tt
  1401.4352}}]\relax
\mciteBstWouldAddEndPuncttrue
\mciteSetBstMidEndSepPunct{\mcitedefaultmidpunct}
{\mcitedefaultendpunct}{\mcitedefaultseppunct}\relax
\EndOfBibitem
\bibitem{Gross:1969jf}
D.~J. Gross and C.~H. Llewellyn~Smith, \emph{{High-energy neutrino--nucleon
  scattering, current algebra and partons}},
  \href{http://dx.doi.org/10.1016/0550-3213(69)90213-2}{\emph{Nucl. Phys.} {\bf
  B14} (1969) 337--347}\relax
\mciteBstWouldAddEndPuncttrue
\mciteSetBstMidEndSepPunct{\mcitedefaultmidpunct}
{\mcitedefaultendpunct}{\mcitedefaultseppunct}\relax
\EndOfBibitem
\bibitem{Baikov:2010iw}
P.~A. Baikov, K.~G. Chetyrkin and J.~H. K{\"u}hn, \emph{{Adler Function, DIS
  sum rules and Crewther Relations}},
  \href{http://dx.doi.org/10.1016/j.nuclphysbps.2010.08.049}{\emph{Nucl. Phys.
  Proc. Suppl.} {\bf 205-206} (2010) 237--241},
  [\href{https://arxiv.org/abs/1007.0478}{{\tt 1007.0478}}]\relax
\mciteBstWouldAddEndPuncttrue
\mciteSetBstMidEndSepPunct{\mcitedefaultmidpunct}
{\mcitedefaultendpunct}{\mcitedefaultseppunct}\relax
\EndOfBibitem
\bibitem{Baikov:2012zn}
P.~A. Baikov, K.~G. Chetyrkin, J.~H. K{\"u}hn and J.~Rittinger, \emph{{Adler
  Function, Sum Rules and Crewther Relation of Order $O(\alpha_s^4)$: the
  Singlet Case}},
  \href{http://dx.doi.org/10.1016/j.physletb.2012.06.052}{\emph{Phys. Lett.}
  {\bf B714} (2012) 62--65}, [\href{https://arxiv.org/abs/1206.1288}{{\tt
  1206.1288}}]\relax
\mciteBstWouldAddEndPuncttrue
\mciteSetBstMidEndSepPunct{\mcitedefaultmidpunct}
{\mcitedefaultendpunct}{\mcitedefaultseppunct}\relax
\EndOfBibitem
\bibitem{Barone:2001sp}
V.~Barone, A.~Drago and P.~G. Ratcliffe, \emph{{Transverse polarisation of
  quarks in hadrons}},
  \href{http://dx.doi.org/10.1016/S0370-1573(01)00051-5}{\emph{Phys. Rept.}
  {\bf 359} (2002) 1--168}, [\href{https://arxiv.org/abs/hep-ph/0104283}{{\tt
  hep-ph/0104283}}]\relax
\mciteBstWouldAddEndPuncttrue
\mciteSetBstMidEndSepPunct{\mcitedefaultmidpunct}
{\mcitedefaultendpunct}{\mcitedefaultseppunct}\relax
\EndOfBibitem
\bibitem{Ralston:1979ys}
J.~P. Ralston and D.~E. Soper, \emph{{Production of Dimuons from High-Energy
  Polarized Proton Proton Collisions}},
  \href{http://dx.doi.org/10.1016/0550-3213(79)90082-8}{\emph{Nucl. Phys.} {\bf
  B152} (1979) 109}\relax
\mciteBstWouldAddEndPuncttrue
\mciteSetBstMidEndSepPunct{\mcitedefaultmidpunct}
{\mcitedefaultendpunct}{\mcitedefaultseppunct}\relax
\EndOfBibitem
\bibitem{Jaffe:1991kp}
R.~L. Jaffe and X.-D. Ji, \emph{{Chiral odd parton distributions and polarized
  Drell-Yan}}, \href{http://dx.doi.org/10.1103/PhysRevLett.67.552}{\emph{Phys.
  Rev. Lett.} {\bf 67} (1991) 552--555}\relax
\mciteBstWouldAddEndPuncttrue
\mciteSetBstMidEndSepPunct{\mcitedefaultmidpunct}
{\mcitedefaultendpunct}{\mcitedefaultseppunct}\relax
\EndOfBibitem
\bibitem{Jaffe:1991ra}
R.~L. Jaffe and X.-D. Ji, \emph{{Chiral odd parton distributions and Drell-Yan
  processes}},
  \href{http://dx.doi.org/10.1016/0550-3213(92)90110-W}{\emph{Nucl. Phys.} {\bf
  B375} (1992) 527--560}\relax
\mciteBstWouldAddEndPuncttrue
\mciteSetBstMidEndSepPunct{\mcitedefaultmidpunct}
{\mcitedefaultendpunct}{\mcitedefaultseppunct}\relax
\EndOfBibitem
\bibitem{Cortes:1991ja}
J.~L. Cortes, B.~Pire and J.~P. Ralston, \emph{{Measuring the transverse
  polarization of quarks in the proton}},
  \href{http://dx.doi.org/10.1007/BF01565099}{\emph{Z. Phys.} {\bf C55} (1992)
  409--416}\relax
\mciteBstWouldAddEndPuncttrue
\mciteSetBstMidEndSepPunct{\mcitedefaultmidpunct}
{\mcitedefaultendpunct}{\mcitedefaultseppunct}\relax
\EndOfBibitem
\bibitem{Artru:1989zv}
X.~Artru and M.~Mekhfi, \emph{{Transversely Polarized Parton Densities, their
  Evolution and their Measurement}},
  \href{http://dx.doi.org/10.1007/BF01556280}{\emph{Z. Phys.} {\bf C45} (1990)
  669}\relax
\mciteBstWouldAddEndPuncttrue
\mciteSetBstMidEndSepPunct{\mcitedefaultmidpunct}
{\mcitedefaultendpunct}{\mcitedefaultseppunct}\relax
\EndOfBibitem
\bibitem{Collins:1992kk}
J.~C. Collins, \emph{{Fragmentation of transversely polarized quarks probed in
  transverse momentum distributions}},
  \href{http://dx.doi.org/10.1016/0550-3213(93)90262-N}{\emph{Nucl. Phys.} {\bf
  B396} (1993) 161--182}, [\href{https://arxiv.org/abs/hep-ph/9208213}{{\tt
  hep-ph/9208213}}]\relax
\mciteBstWouldAddEndPuncttrue
\mciteSetBstMidEndSepPunct{\mcitedefaultmidpunct}
{\mcitedefaultendpunct}{\mcitedefaultseppunct}\relax
\EndOfBibitem
\bibitem{Jaffe:1993xb}
R.~L. Jaffe and X.-D. Ji, \emph{{Novel quark fragmentation functions and the
  nucleon's transversity distribution}},
  \href{http://dx.doi.org/10.1103/PhysRevLett.71.2547}{\emph{Phys. Rev. Lett.}
  {\bf 71} (1993) 2547--2550},
  [\href{https://arxiv.org/abs/hep-ph/9307329}{{\tt hep-ph/9307329}}]\relax
\mciteBstWouldAddEndPuncttrue
\mciteSetBstMidEndSepPunct{\mcitedefaultmidpunct}
{\mcitedefaultendpunct}{\mcitedefaultseppunct}\relax
\EndOfBibitem
\bibitem{Tangerman:1994bb}
R.~D. Tangerman and P.~J. Mulders, \emph{{Polarized twist-three distributions
  $g_T$ and $h_L$ and the role of intrinsic transverse momentum}},
  \href{https://arxiv.org/abs/hep-ph/9408305}{{\tt hep-ph/9408305}}\relax
\mciteBstWouldAddEndPuncttrue
\mciteSetBstMidEndSepPunct{\mcitedefaultmidpunct}
{\mcitedefaultendpunct}{\mcitedefaultseppunct}\relax
\EndOfBibitem
\bibitem{Boer:1997nt}
D.~Boer and P.~J. Mulders, \emph{{Time reversal odd distribution functions in
  leptoproduction}},
  \href{http://dx.doi.org/10.1103/PhysRevD.57.5780}{\emph{Phys. Rev.} {\bf D57}
  (1998) 5780--5786}, [\href{https://arxiv.org/abs/hep-ph/9711485}{{\tt
  hep-ph/9711485}}]\relax
\mciteBstWouldAddEndPuncttrue
\mciteSetBstMidEndSepPunct{\mcitedefaultmidpunct}
{\mcitedefaultendpunct}{\mcitedefaultseppunct}\relax
\EndOfBibitem
\end{mcitethebibliography}\endgroup

\end{document}